\documentclass[12pt]{article}

\usepackage[margin=1in,nohead,dvips,a4paper]{geometry}
\usepackage{times,pstricks,pst-node}
\usepackage{amssymb}
\usepackage[small,center]{caption}

\newtheorem{theorem}{Theorem}
\newtheorem{lemma}{Lemma}

\makeatletter
\DeclareRobustCommand{\qed}{%
  \ifmmode 
  \else \leavevmode\unskip\penalty9999 \hbox{}\nobreak\hfill
  \fi
  \quad\hbox{\qedsymbol}}
\newcommand{\openbox}{\leavevmode
  \hbox to.77778em{%
  \hfil\vrule
  \vbox to.675em{\hrule width.6em\vfil\hrule}%
  \vrule\hfil}}
\newcommand{\qedsymbol}{\openbox}
\newenvironment{proof}[1][\proofname]{\par
  \normalfont
  \topsep6\p@\@plus6\p@ \trivlist
  \item[\hskip\labelsep\itshape
    #1.]\ignorespaces
}{%
  \qed\endtrivlist
}
\newcommand{\proofname}{Proof}
\makeatother

\psset{linewidth=.5pt,dash=4pt 4pt,unit=.25in,arrowsize=2pt 3}

\def\arrowLine(#1,#2)(#3,#4){%
  \pcline(#1,#2)(#3,#4)%
  \lput{:U}{
    \pspicture(0,0)(0,0)
      \psline[arrows=->](2.3pt,0)(2.4pt,0)
    \endpspicture
  }
}

\def\vertexA #1{%
  \pspicture(0,0)(2,2)
    \arrowLine(0,1)(1,1)
    \arrowLine(1,1)(1,2)
    \arrowLine(2,1)(1,1)
    \arrowLine(1,1)(1,0)
    \psdot(1,1)
    \rput(0.6,0.6){#1}
  \endpspicture
}

\def\vertexB #1{%
  \pspicture(0,0)(2,2)
    \arrowLine(1,1)(0,1)
    \arrowLine(1,2)(1,1)
    \arrowLine(1,1)(2,1)
    \arrowLine(1,0)(1,1)
    \psdot(1,1)    
    \rput(0.6,0.6){#1}
  \endpspicture
}

\def\vertexC #1{%
  \pspicture(0,0)(2,2)
    \arrowLine(0,1)(1,1)
    \arrowLine(1,1)(1,2)
    \arrowLine(1,1)(2,1)
    \arrowLine(1,0)(1,1)
    \psdot(1,1)    
    \rput(0.6,0.6){#1}
  \endpspicture
}

\def\vertexD #1{%
  \pspicture(0,0)(2,2)
    \arrowLine(1,1)(0,1)
    \arrowLine(1,2)(1,1)
    \arrowLine(2,1)(1,1)
    \arrowLine(1,1)(1,0)
    \psdot(1,1)    
    \rput(0.6,0.6){#1}
  \endpspicture
}

\def\vertexE #1{%
  \pspicture(0,0)(2,2)
    \arrowLine(1,1)(0,1)
    \arrowLine(1,1)(1,2)
    \arrowLine(2,1)(1,1)
    \arrowLine(1,0)(1,1)
    \psdot(1,1)    
    \rput(0.6,0.6){#1}
  \endpspicture
}

\def\vertexF #1{%
  \pspicture(0,0)(2,2)
    \arrowLine(0,1)(1,1)
    \arrowLine(1,2)(1,1)
    \arrowLine(1,1)(2,1)
    \arrowLine(1,1)(1,0)
    \psdot(1,1)    
    \rput(0.6,0.6){#1}
  \endpspicture
}

\def\bm #1{\mbox{\boldmath $#1$}}

\mathchardef\slantPhi="108
\mathchardef\slantSigma="106

\def\inv {\mathop{\mathrm{inv}}}
\def\HT {\mathrm{HT}}
\def\rme {\mathrm e}
\def\rmi {\mathrm i}

\begin{document}

\title{Enumerations of half-turn symmetric alternating-sign
matrices of odd order}
\author{A.~V.~Razumov, Yu.~G.~Stroganov\\
\small \it Institute for High Energy Physics\\[-.5em]
\small \it 142280 Protvino, Moscow region, Russia}
\date{}

\maketitle

\begin{abstract}
It was shown by Kuperberg that the partition function of the square-ice
model related to half-turn symmetric alternating-sign matrices of even
order is the product of two similar factors. We propose a square-ice model
whose states are in bijection with half-turn symmetric alternating-sign
matrices of odd order. The partition function of the model is expressed via
the above mentioned factors. The contributions to the partition function
of the states corresponding to the alternating-sign matrices having~$1$
or~$-1$ as the central entry are found and the related enumerations are
obtained.
\end{abstract}

\section{Introduction}

An alternating-sign matrix is a matrix with entries $1$, $0$, and $-1$ such
that the $1$ and $-1$ entries alternate in each column and each row and
such that the first and last nonzero entries in each row and column are
$1$. Starting  from the famous conjectures by Mills, Robbins and Rumsey
\cite{MilRobRum82, MilRobRum83} a lot of enumeration and equinumeration
results on alternating-sign matrices and their various subclasses were
obtained. Most of the results were proved using bijections between matrices
and states of different variants of the statistical square-ice model. For
the first time such a method to solve enumeration problems was used by
Kuperberg~\cite{Kup96}, see also the rich in results paper~\cite{Kup02}.

The present paper is devoted to the study of enumerations of the half-turn
symmetric alternating-sign matrices of odd order on the base of the
corresponding square-ice model.

In Section 2 we recall the definition of the square-ice model with the
domain wall boundary conditions. The states of this model are in bijection
with the alternating sign-matrices. The necessary properties of the
partition function of the model are established.

In Section 3 we discuss first the square-ice model related to the half-turn
symmetric alter\-na\-ting-sign matrices of even order proposed by
Kuperberg \cite{Kup02}. Then a square-ice model whose states are in
bijection with the half-turn symmetric alternating-sign matrices of odd
order is introduced. We prove that the partition sum of this model is
expressed via the partition function of the square-ice model with domain
wall boundary condition and the partition function of the square-ice model
related to half-turn symmetric alternating-sign matrices of even order
(Theorem \ref{t:1}). It appears that one can separate the contributions to
the partition function of the states corresponding to the alternating-sign
matrices having~$1$ and~$-1$ as the central entry (Theorem \ref{t:2}).

In Section 4 we consider an important special case of the overall
parameter of the model which allow to prove, in particular, the
enumeration conjectures by Robbibs \cite{Rob00} on half-turn symmetric
alternating-sign matrices of odd order. It is interesting that in this case
there is a factorised determinant representation of the partition function
(Theorem \ref{t:3}).

In section 5 we relate the enumerations of half-turn symmetric
alternating-sign matrices of odd order with the enumerations of general
alternating-sign matrices and half-turn symmetric alternating-sign matrices
of even order. In particular, we find the explicit separate refined
enumerations of the half-turn symmetric alternating-sign matrices of odd
order having~$1$ and~$-1$ in the center of matrix.

\section{Square-ice model with domain wall boundary}

\subsection{Definition of the model}

\begin{figure}[ht]
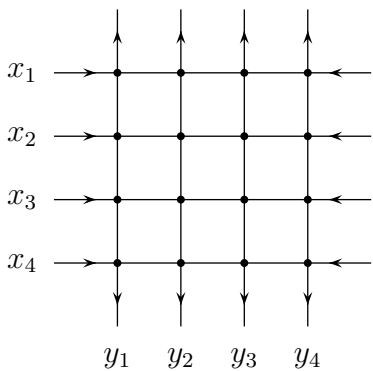

\[
\psset{unit=2em}
\pspicture[.4](-1,-.7)(5,5)
  \arrowLine(0,1)(1,1)
  \arrowLine(0,2)(1,2)
  \arrowLine(0,3)(1,3)
  \arrowLine(0,4)(1,4)
  \arrowLine(5,1)(4,1)
  \arrowLine(5,2)(4,2)
  \arrowLine(5,3)(4,3)
  \arrowLine(5,4)(4,4)
  \arrowLine(1,1)(1,0)
  \arrowLine(2,1)(2,0)
  \arrowLine(3,1)(3,0)
  \arrowLine(4,1)(4,0)
  \arrowLine(1,4)(1,5)
  \arrowLine(2,4)(2,5)
  \arrowLine(3,4)(3,5)
  \arrowLine(4,4)(4,5)
  \psline(1,1)(4,1)
  \psline(1,2)(4,2)
  \psline(1,3)(4,3)
  \psline(1,4)(4,4)
  \psline(1,1)(1,4)
  \psline(2,1)(2,4)
  \psline(3,1)(3,4)
  \psline(4,1)(4,4)
  \psdot(1,1) \psdot(2,1) \psdot(3,1) \psdot(4,1)
  \psdot(1,2) \psdot(2,2) \psdot(3,2) \psdot(4,2)
  \psdot(1,3) \psdot(2,3) \psdot(3,3) \psdot(4,3)
  \psdot(1,4) \psdot(2,4) \psdot(3,4) \psdot(4,4)  
  \rput(-.5,4){$x_1$}
  \rput(-.5,3){$x_2$}
  \rput(-.5,2){$x_3$}
  \rput(-.5,1){$x_4$}
  \rput(1,-.5){$y_1$}
  \rput(2,-.5){$y_2$}
  \rput(3,-.5){$y_3$}
  \rput(4,-.5){$y_4$}
\endpspicture
\]
\caption{Square ice with a domain-wall boundary}
\label{f:dw}
\end{figure}

The method used by Kuperberg to prove the alternating-sign matrix
conjecture is based on the bijection between the states of the square-ice
model with the domain wall boundary conditions and alternating-sign
matrices. To define the state space of the square ice model we consider a
subset of vertices and edges of a square grid, such that each internal
vertex is tetravalent and each boundary vertex is univalent. A state of a
corresponding square ice model is determined by orienting the edges in such
a way that two edges enter and leave every tetravalent vertex. The domain
wall boundary conditions \cite{Kor82} fix the orientation of the edges
belonging to univalent vertices in accordance with the pattern given in
Figure~\ref{f:dw}.
The labels $x_i$ and $y_i$ are the spectral parameters which will be used
to define the partition function of the model.

\begin{figure}[ht]
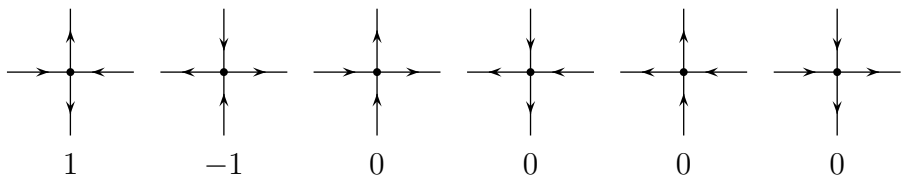

\[
\psset{unit=2em}
\begin{array}{cccccc}
\vertexA{} & \vertexB{} & \vertexC{} & \vertexD{} & \vertexE{} & \vertexF{}
\\
1 & -1 & 0 & 0 & 0 & 0
\end{array}
\]
\caption{The correspondence between the square ice vertices
and the entries of alternating-sign matrices}
\label{f:dwasm}
\end{figure}

If we replace each tetravalent vertex of a state of the square ice
with a domain wall boundary condition by a number according to
Figure~\ref{f:dwasm} we will obtain an alternating-sign matrix.
It is not difficult to check that in this way we come to the bijection
between the states and the alternating-sign matrices \cite{RobRum86,
ElkKupLarPro92}.

\begin{figure}[ht]
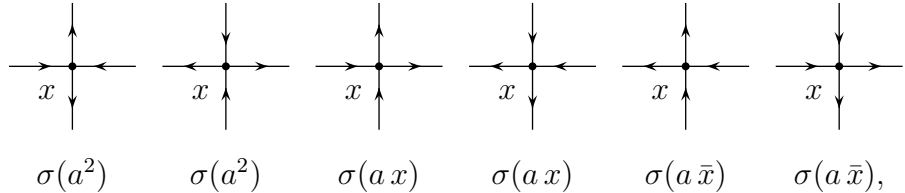

\[
\psset{unit=2em}
\begin{array}{cccccc}
\vertexA{$x$} & \vertexB{$x$} & \vertexC{$x$} & \vertexD{$x$} &
\vertexE{$x$} & \vertexF{$x$} \\[.5em]
\sigma(a^2) & \sigma(a^2) & \sigma(a \, x) & \sigma(a \, x) & \sigma(a \,
\bar x) & \sigma(a \, \bar x),
\end{array}
\]
\caption{The weights of the vertices}
\label{f:wghts}
\end{figure}

The partition function of the square ice with a domain wall
boundary is the sum of the weights of all possible states of the model. The
weight of a state is the product of the weights of all tetravalent
vertices. To define them, we associate spectral parameters $x_i$
with the vertical lines of the grid and spectral parameters $y_i$ with the
horizontal ones (see Figure~\ref{f:dw}). A vertex at the intersection of
the line with the spectral parameters $x_i$ and $y_j$ is supplied with the
spectral parameter $x_i \bar y_j$, where we use the notation $\bar x =
x^{-1}$ introduced by Kuperberg. After that we define the weights of the
vertices as it is given in Figure~\ref{f:wghts},
where $a$ is a parameter common for all vertices and we use the convenient
abbreviations $\sigma(x) = x - \bar x$ also introduced by Kuperberg.

A graph, similar to one given in Figure \ref{f:dw}, with labeled vertices
denotes the corresponding partition function. Here the summation over
all possible orientations of internal edges is implied. If we have
unoriented boundary edges, then the graph represents the set of the
quantities corresponding to their possible orientations. It is convenient
to make the formalism invariant with respect to rotations. To this end we
allow a vertex label to be positioned in any quadrant related to the
vertex. The value of the corresponding spectral parameter is equal to $x_i
\bar y_j$ if the label is placed into the quadrant swept by the line with
the spectral parameter $x_i$ when it is rotated anticlockwise to the line
with the spectral parameter $y_j$. This rule implies that we can move a
label $x$ from one quadrant to an adjacent one, changing it to $\bar x$.

\begin{figure}[ht]
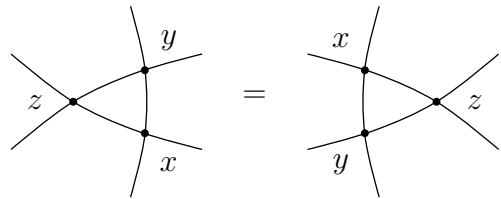
 
\[
\pspicture[.5](0,0)(4,4)
\pscurve(0,1)(1.3,2.0)(2.6,2.6)(4,3)
\pscurve(0,3)(1.3,2.0)(2.6,1.4)(4,1)
\pscurve(2.5,4)(2.8,2.7)(2.8,1.3)(2.5,0)
\rput(.5,2){$z$}
\rput(3.3,3.3){$y$}
\rput(3.3,.7){$x$}
\psdot(1.3,2)
\psdot(2.8,1.34)
\psdot(2.8,2.66)
\endpspicture
\quad = \quad
\pspicture[.5](0,0)(4,4)
\pscurve(4,1)(2.7,2.0)(1.4,2.6)(0,3)
\pscurve(4,3)(2.7,2.0)(1.4,1.4)(0,1)
\pscurve(1.5,4)(1.2,2.7)(1.2,1.3)(1.5,0)
\rput(3.5,2){$z$}
\rput(.7,3.3){$x$}
\rput(.7,.7){$y$}
\psdot(2.7,2)
\psdot(1.2,1.34)
\psdot(1.2,2.66)
\endpspicture
\]
\caption{The Yang--Baxter equation}
\label{f:yb}
\end{figure}

As an example we give the graph corresponding to the well-known
Yang--Baxter equation, see Figure \ref{f:yb}. This equation is satisfied if
$xyz = a$.
Note that we use the parametrization of the weights convenient for
consideration of combinatorial problems proposed by Kuperberg. 

\subsection{Two lemmas on permutations}

\begin{figure}[ht]
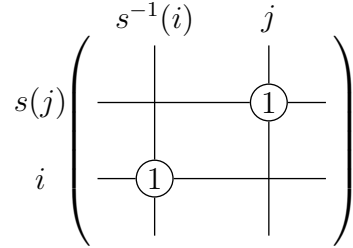

\[
\psset{unit=1.2em}
\pspicture(0,0)(9,6)
\rput(0.5,1.5){$i$}
\rput(0.5,3.5){$s(j)$}
\rput(1.5,2.5){$\left( \vrule height 2.5\psunit depth 2.5\psunit width 0 pt
\right.$}
\rput(8.5,2.5){$\left. \vrule height 2.5\psunit depth 2.5\psunit width 0 pt
\right)$}
\rput(3.5,5.7){$s^{-1}(i)$}
\rput(6.5,5.7){$j$}
\rput(6.5,3.5){$1$}
\pscircle(6.5,3.5){.5}
\rput(3.5,1.5){$1$}
\pscircle(3.5,1.5){.5}
\psline(2,3.5)(6,3.5)
\psline(7,3.5)(8,3.5)
\psline(2,1.5)(3,1.5)
\psline(4,1.5)(8,1.5)
\psline(3.5,0)(3.5,1)
\psline(3.5,2)(3.5,5)
\psline(6.5,0)(6.5,3)
\psline(6.5,4)(6.5,5)
\endpspicture
\]
\caption{Positions of 1 entries in rows and columns of a permutation
matrix}
\label{f:pou}
\end{figure}

A simplest example of an alternating-sign matrix is a permutation matrix,
which is defined as a matrix which can be created by rearranging the
rows and columns of an identity matrix. For any $n \times n$ permutation
matrix $\slantSigma$ one can write
\[
(\slantSigma)_{ij} = \delta_{i s(j)},
\]
where $s$ is an appropriate unique element of the symmetric group $S_n$. It
is clear that if we go along the column of $\slantSigma$ with the number
$j$ we meet the 1 entry at the row with the number $s(j)$, and if we go
along the row of $\slantSigma$ with the number $i$ we meet the 1 entry at
the column with the number $s^{-1}(i)$, see Figure~\ref{f:pou}.

Recall that a pair $(i, j)$ of integers such that $1 \le i < j \le n$ is
said to be an inversion of a permutation $s \in S_n$ if $s(i) > s(j)$. The
number of inversions of $s$ is denoted as $\inv(s)$. It is clear that the
minimal number of inversions of an element of $S_n$ is zero, and the
maximal one is $n(n - 1)/2$. Introduce the generating
function
\[
\slantPhi(n; z) = \sum_{s \in S_n} z^{\inv(s)} = \sum_{k=0}^{n(n - 1)/2}
I(n; k) z^k,
\]
where $I(n; k)$ is the number of the elements of $S_n$ with $k$ inversions.
The following result is well known.

\begin{lemma} \label{l:1}
For any positive integer $n$ one has
\begin{equation}
\slantPhi(n; z) = (1 + z)(1 + z + z^2) \cdots (1 + z + \cdots + z^{n-1}).
\label{e:1}
\end{equation}
\end{lemma}

\begin{proof}
The very form of the statement of the lemma suggests to prove it by
induction. Represent an element $s$ of $S_n$ as the word $s(1)s(2)
\ldots s(n)$. The words representing the elements of $S_{n+1}$ can be
created by inserting the letter `$n+1\!$' into the words representing the
elements of $S_n$. As the result we have $n+1$ possibilities. First, let
the letter `$n+1\!$' is at the last position of the resulting word. The
length of the obtained element of $S_{n+1}$ is equal to the length of the
initial element of $S_n$. This possibility gives a contribution to
$\slantPhi(n+1; z)$  coinciding with $\slantPhi(n; z)$. Let now the letter
`$n+1\!$' is at the next to last position. The length of the obtained
permutation is one more than the length of the initial one. This gives a
contribution $\slantPhi(n; z) z$. Exhausting all the possibilities we
obtain the equality
\[
\slantPhi(n+1; z) = \slantPhi(n; z) (1 + z + \cdots + z^n).
\]
Taking into account that $\slantPhi(1; z) = 1$, we come to the statement of
the lemma.
\end{proof}

\begin{lemma} \label{l:2}
For a state of square ice corresponding to the permutation matrix
associated with a permutation $s \in S_n$, the numbers of the vertices of
third type and of forth type are equal to $\inv(s)$ and the numbers of the
vertices of fifth type and of sixth type are equal to $n(n-1)/2 - \inv(s)$.
\end{lemma}

\begin{proof}
Consider a state of square ice corresponding to a permutation
matrix $\slantSigma$ associated with a permutation $s \in S_n$. Note that
if we walk along a row or a column of the picture representing the state
under consideration, then passing a vertex of the first or the second type
we change the orientation of the edge. Taking into account that the
vertices
of the first or the second type are situated where one finds entry 1 in the
corresponding permutation matrix one can make the following conclusions. A
0 entry in the permutation matrix corresponds to a vertex of third type, if
a 1 entry is below it and a 1 entry is from the right of it; a 0 entry in
the permutation matrix corresponds to a vertex of forth type, if a 1 entry
is above it and a 1 entry is from the left of it, see Figure~\ref{f:lvp}.
\begin{figure}[ht]
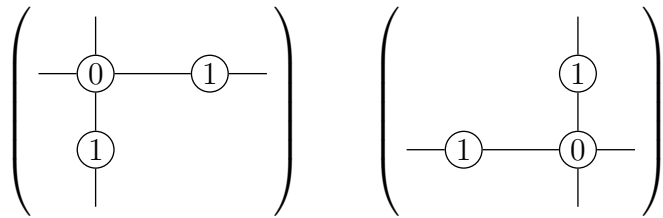

\[
\psset{unit=1.2em}
\pspicture(0,0)(8,5)
\rput(0.5,2.5){$\left( \vrule height 2.5\psunit depth 2.5\psunit width 0 pt
\right.$}
\rput(7.5,2.5){$\left. \vrule height 2.5\psunit depth 2.5\psunit width 0 pt
\right)$}
\rput(2.5,3.5){$0$}
\pscircle(2.5,3.5){.5}
\rput(5.5,3.5){$1$}
\pscircle(5.5,3.5){.5}
\rput(2.5,1.5){$1$}
\pscircle(2.5,1.5){.5}
\psline(1,3.5)(2,3.5)
\psline(3,3.5)(5,3.5)
\psline(6,3.5)(7,3.5)
\psline(2.5,0)(2.5,1)
\psline(2.5,2)(2.5,3)
\psline(2.5,4)(2.5,5)
\endpspicture
\qquad
\pspicture(0,0)(8,5)
\rput(0.5,2.5){$\left( \vrule height 2.5\psunit depth 2.5\psunit width 0 pt
\right.$}
\rput(7.5,2.5){$\left. \vrule height 2.5\psunit depth 2.5\psunit width 0 pt
\right)$}
\rput(5.5,1.5){$0$}
\pscircle(5.5,1.5){.5}
\rput(5.5,3.5){$1$}
\pscircle(5.5,3.5){.5}
\rput(2.5,1.5){$1$}
\pscircle(2.5,1.5){.5}
\psline(1,1.5)(2,1.5)
\psline(3,1.5)(5,1.5)
\psline(6,1.5)(7,1.5)
\psline(5.5,0)(5.5,1)
\psline(5.5,2)(5.5,3)
\psline(5.5,4)(5.5,5)
\endpspicture
\]
\caption{The disposition of the vertices of third and forth types}
\label{f:lvp}
\end{figure}
Similarly, a 0 entry in the permutation matrix corresponds to a vertex of
fifth type, if a 1 entry is below it and a 1 entry is from the left of
it; a 0 entry in the permutation matrix corresponds to a vertex of sixth
type, if a 1 entry is above it and a 1 entry is from the right of
it, see Figure~\ref{f:rvp}.
\begin{figure}[ht]
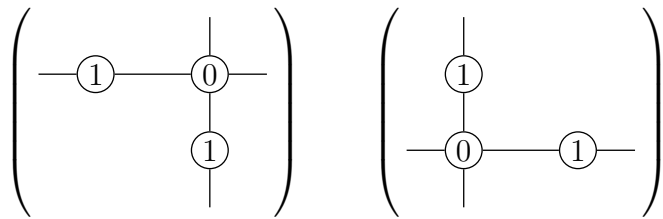

\[
\psset{unit=1.2em}
\pspicture(0,0)(8,5)
\rput(0.5,2.5){$\left( \vrule height 2.5\psunit depth 2.5\psunit width 0 pt
\right.$}
\rput(7.5,2.5){$\left. \vrule height 2.5\psunit depth 2.5\psunit width 0 pt
\right)$}
\rput(2.5,3.5){$1$}
\pscircle(2.5,3.5){.5}
\rput(5.5,3.5){$0$}
\pscircle(5.5,3.5){.5}
\rput(5.5,1.5){$1$}
\pscircle(5.5,1.5){.5}
\psline(1,3.5)(2,3.5)
\psline(3,3.5)(5,3.5)
\psline(6,3.5)(7,3.5)
\psline(5.5,0)(5.5,1)
\psline(5.5,2)(5.5,3)
\psline(5.5,4)(5.5,5)
\endpspicture
\qquad
\pspicture(0,0)(8,5)
\rput(0.5,2.5){$\left( \vrule height 2.5\psunit depth 2.5\psunit width 0 pt
\right.$}
\rput(7.5,2.5){$\left. \vrule height 2.5\psunit depth 2.5\psunit width 0 pt
\right)$}
\rput(5.5,1.5){$1$}
\pscircle(5.5,1.5){.5}
\rput(2.5,3.5){$1$}
\pscircle(2.5,3.5){.5}
\rput(2.5,1.5){$0$}
\pscircle(2.5,1.5){.5}
\psline(1,1.5)(2,1.5)
\psline(3,1.5)(5,1.5)
\psline(6,1.5)(7,1.5)
\psline(2.5,0)(2.5,1)
\psline(2.5,2)(2.5,3)
\psline(2.5,4)(2.5,5)
\endpspicture
\]
\caption{The disposition of the vertices of fifth and sixth types}
\label{f:rvp}
\end{figure}

Let now $(i, j)$ be an inversion of $s$. Consider the item at the
intersection of the row with the number $s(i)$ and the column with the
number $j$ of the matrix $\slantSigma$. Since $(i, j)$ is an inversion, we
have the situation depicted in Figure~\ref{f:inv}.
\begin{figure}[ht]
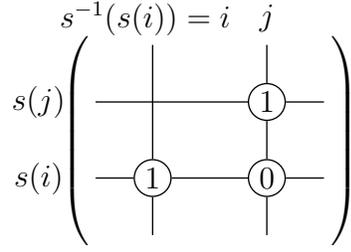

\[
\psset{unit=1.2em}
\pspicture(0,0)(9,6)
\rput(0.5,1.5){$s(i)$}
\rput(0.5,3.5){$s(j)$}
\rput(1.5,2.5){$\left( \vrule height 2.5\psunit depth 2.5\psunit width 0 pt
\right.$}
\rput(8.5,2.5){$\left. \vrule height 2.5\psunit depth 2.5\psunit width 0 pt
\right)$}
\rput(3.3,5.7){$s^{-1}(s(i)) = i$}
\rput(6.5,5.7){$j$}
\rput(6.5,3.5){$1$}
\pscircle(6.5,3.5){.5}
\rput(3.5,1.5){$1$}
\pscircle(3.5,1.5){.5}
\rput(6.5,1.5){$0$}
\pscircle(6.5,1.5){.5}

\psline(2,3.5)(6,3.5)
\psline(7,3.5)(8,3.5)
\psline(2,1.5)(3,1.5)
\psline(4,1.5)(6,1.5)
\psline(7,1.5)(8,1.5)
\psline(3.5,0)(3.5,1)
\psline(3.5,2)(3.5,5)
\psline(6.5,0)(6.5,1)
\psline(6.5,2)(6.5,3)
\psline(6.5,4)(6.5,5)
\endpspicture
\]
\caption{The correspondence beetwen the inversions and the vertices of the
forth type}
\label{f:inv}
\end{figure}
It is clear that the state of square ice have at the considered position a
vertex of forth type. It is not difficult to get convinced that we have a
bijection between the inversions of $s$ and the vertices of the forth type
of the state. Hence, the number of the vertices of forth type is equal to
$\inv(s)$. Analysing Figure~\ref{f:lvp}, one sees that we have equal
numbers of the vertices of the third type and of the forth type.

Further, Figure~\ref{f:lvp} shows that we have equal numbers of the
vertices of the fifth type and of the sixth type. Taking into account that
the total number of the vertices of the first type and of the second type
is $n$, we see that the last statement of the lemma is true.
\end{proof}

\subsection{Leading terms of the partition function}

Denote the partition function of the square-ice model with a domain wall
boundary condition by $Z(n; \bm x, \bm y)$. Here $n$ is the size of the
square ice, $\bm x$ and $\bm y$ are vectors constructed from the
spectral parameters,
\[
\bm x = (x_1, \ldots, x_n), \qquad \bm y = (x_1, \ldots, x_n).
\]
It is clear that the partition function $Z(n; \bm x, \bm y)$ is a Laurent
polynomial in the variables $x_i$ and $y_i$. Define the modified partition
function
\[
\widetilde Z(n; \bm x, \bm y) = \left[ \prod_{i=1}^n x_i^{n-1}
y_i^{n-1} \right] Z(n; \bm x, \bm y).
\]
The next lemma describes some properties of the modified partition function
which will be needed below.

\begin{lemma} \label{l:3} 
The  modified partition function $\widetilde Z(n; \bm x, \bm y)$ has the
following properties.
\begin{itemize}
\item[\rm (a)]
The function $\widetilde Z(n; \bm x, \bm y)$ is symmetric separately in the
variables $x_1, \ldots, x_n$ and in the variables $y_1, \ldots, y_n$.
\item[\rm (b)]
The function $\widetilde Z(n; \bm x, \bm y)$ is a homogeneous polynomial in
the variables $x_i$ and $y_i$ of total degree $2n (n-1)$. For each fixed $i
= 1, \ldots, n$ it is a polynomial in $x_i^2$ of degree $n-1$ and a
polynomial in $y_i^2$ of degree $n-1$.
\item[\rm (c)]
If~$y_n = a x_n$,
then
\begin{equation}
\widetilde Z(n; \bm x, \bm y) = \sigma(a^2) \prod_{i=1}^{n-1} \left[ (a
y_i^2 - \bar a x_n^2) (a y_n^2 - \bar a x_i^2) \right] \widetilde Z(n-1;
\bm x \smallsetminus x_n, \bm y \smallsetminus y_n), \label{e:2}
\end{equation}
where $\bm x \smallsetminus x_n = (x_1, \ldots, x_{n-1})$ and $\bm y
\smallsetminus y_n = (y_1, \ldots, y_{n-1})$.
\end{itemize}
\end{lemma}

\begin{proof} 
Statement (a) of the lemma follows from the symmetricity of the partition
function $Z(n; \bm x, \bm y)$. Statement (b) is actually evident.
The recursive relation (\ref{e:2}) follows from the corresponding recursive
relation for the partition function $Z(n; \bm x, \bm y)$ which states that
if $y_n = a x_n$, then
\begin{equation}
Z(n; \bm x, \bm y) = \sigma(a^2) \prod_{i=1}^{n-1} \left[ \sigma(a \, \bar
x_n \, y_i) \sigma(a \, \bar x_i \, y_n) \right] Z(n-1; \bm x
\smallsetminus x_n, \bm y \smallsetminus y_n). \label{e:3}
\end{equation}
The proof of the mentioned properties of the partition function $Z(n; \bm
x, \bm y)$ can be found, for example, in paper \cite{Kup02}.
\end{proof}

In calculating the modified partition function $\widetilde Z(n; \bm x, \bm
y)$ it is convenient to multiply the weight of a vertex at the intersection
of the horizontal line with the spectral parameter $x_i$ and the vertical
line with the spectral parameter $y_j$  corresponding to a 0 entry of the
alternating-sign matrix by $x_i y_j$. Here the factor $x_i y_j$ is
extracted from the factor entering the definition of the modified partition
function. Hence, we assume that a 0 entry corresponding to a vertex of
third or forth type gives the contribution $a x_i^2 - \bar a y_j^2$ to the
corresponding term of the modified partition function, and a 0 entry
corresponding to a vertex of fifth or sixth type gives the contribution $a
y_j^2 - \bar a x_i^2$. 

Singling out from $\widetilde Z(n; \bm x, \bm y)$ the term of maximal
degree in the variables $x_i$, we write
\[
\widetilde Z(n; \bm x, \bm y) = \left[ \prod_{i=1}^n x_i^{2(n-1)} \right]
C(n) + \ldots .
\]

\begin{lemma} \label{l:4}
The coefficient $C(n)$ is given by the formula
\begin{equation}
C(n) = \prod_{i=1}^{n} \sigma(a^{2i}). \label{e:4}
\end{equation}
\end{lemma}

\begin{proof}
One can easily get convinced that to find the term of maximal degree in the
variables $x_i$ one should take only the states of the square ice
which have only one 1 entry in each column of the corresponding
alternating-sign matrices. Only in this case all the parameters $y_i$ from
the factor entering the definition of the modified partition function can
be absorbed into weights of vertices. The alternating-sign matrices which
have only one 1 entry in each column are the permutation matrices.
Therefore, from Lemma \ref{l:2} it follows that the contribution to the
term of maximal degree in the variables $x_i$ of the state corresponding to
a permutation $s$ is
\[
\left[ \prod_{i=1}^n x_i^{2(n-1)} \right] \sigma^n(a^2) a^{2 \inv(s)}
(-\bar a)^{n(n-1) - 2 \inv(s)} = \left[ \prod_{i=1}^n x_i^{2(n-1)} \right]
\sigma^n(a^2) a^{-n(n-1)} a^{4 \inv(s)}.
\]
Hence, one obtains
\[
C(n) = \sigma^n(a^2) a^{-n(n-1)} \sum_{s \in S_n} a^{4 \inv(s)}.
\]
Taking into account equality (\ref{e:1}), after some elementary
transformations we come to relation~(\ref{e:4}).
\end{proof}

Let us go further and consider the terms of $\widetilde Z(n; \bm x, \bm y)$
which have the maximal total degree in all the variables $x_i$ except the
variable $x_n$, and do not contain all the variables $y_i$ except the
variable $y_n$. Singling out these term, we write
\[
\widetilde Z(n; \bm x, \bm y) = \left[ \prod_{i=1}^{n-1} x_i^{2(n-1)}
\right] S(n; x_n, y_n) + \ldots .
\]

\begin{lemma}
The polynomial $S(n; x_n, y_n)$ has the form
\begin{equation}
S(n; x_n, y_n) = \left[ \prod_{i=1}^{n-1} \sigma(a^{2i}) \right]
\left( \sigma(a^{2n}) x_n^{2(n-1)} - \sigma(a^{2(n-1)}) x_n^{2(n-2)} y_n^2
\right). \label{e:5}
\end{equation}
\end{lemma}

\begin{proof}
Since we are looking for the term which do not contain all the variable
$y_i$ except the variable $y_n$, we can consider only the states which have
only one 1 entry in each of the first $n-1$ columns. By the definition of
alternating-sign matrix one has only one 1 entry in the last column too.
The first $n-1$ rows of such a matrix give the required degree in the
variables $x_i$. The weight of a vertex in the last row is either
$a y_j^2 - \bar a x_n^2$ or $a x_n^2 - \bar a y_j^2$. One has only one
entry 1 in this row. If this entry is in the last column, we have the
degree $2(n-1)$ in the variable $x_n$ and zero degree in the variable
$x_n$. If the entry 1 is not in the last column we obtain either a
contribution proportional $x_n^{2(n-1)}$ or a contribution proportional to
$x_n^{2(n-2)} y_n^2$ depending on which term we take from the weight of the
last vertex in the row equal to $a x_n - \bar a y_n$. Thus, we conclude
that the polynomial $S(n; x_n, y_n)$ is of the form
\[
S(n; x_n, y_n) = C(n) x_n^{2(n-1)} + D(n) x_n^{2(n-2)} y_n^2.
\]
From recursive relation (\ref{e:2}) one obtains
\[
C(n) + a^2 D(n) = \sigma(a^2) a^{-2(n-1)} C(n-1).
\]
Taking into account equality (\ref{e:4}), one comes to the relation
\[
D(n) = - \left[ \prod_{i=1}^{n-1} \sigma(a^{2i}) \right] \sigma(a^{2(n-1)})
\]
which implies equality (\ref{e:5}).
\end{proof}

\section{Square-ice models related to half-turn symmetric
alter\-na\-ting-sign matrices}

\subsection{Half-turn symmetric alternating-sign matrices}

We say that an alternating-sign matrix $A$ is half-turn symmetric if
\[
(A)_{n-1-i,n-1-j} = (A)_{i,j}.
\]
The $n \times n$ permutation matrix $\Sigma$ associated with a permutation
$s \in S_n$ is half-turn symmetric if and only if
\[
s(n - 1 - i) = n - 1 - s(i), \qquad i = 1, \ldots n.
\]
We denote the set of all permutations associated with $n \times n$
half-turn symmetric permutation matrices by $S^{\mathrm{HT}}_n$ and
introduce the generating function
\[
\slantPhi_{\mathrm{HT}}(n; z) = \sum_{s \in S^{\mathrm{HT}}_n} z^{\inv(s)}
= \sum_{k=0}^{n(n - 1)/2} I_{\mathrm{HT}}(n; k) z^k,
\]
where $I_{\mathrm{HT}}(n; k)$ is the number of the elements of
$S^{\mathrm{HT}}_n$ with $k$ inversions.

\begin{lemma} \label{l:6}
The equalities
\begin{eqnarray}
& \displaystyle \slantPhi_{\mathrm{HT}}(2m+1; z) = \left[ \prod_{i=1}^m (1
+ z^{2i+1}) \right] \slantPhi(m; z^2), \label{e:6} \\[.5em]
& \displaystyle \slantPhi_{\mathrm{HT}}(2m; z) = \left[ \prod_{i=1}^m (1 +
z^{2i-1}) \right] \slantPhi(m; z^2) \label{e:7}
\end{eqnarray}
are valid.
\end{lemma}

\begin{proof} 
Equality (\ref{e:6}) is valid for $m=0$, suppose that it is
valid for some $m = k > 0$. Let $s$ be an arbitrary element of
$S^{\mathrm{HT}}_{2k+1}$. Identify it with the permutation $s'$ of the
alphabet $\{2, 3, \ldots, 2k+2\}$ defined as
\[
s'(i) = s(i-1)+1
\]
and represent $s$ as the word $s'(2)s'(3)\ldots s'(2k+2)$. If we insert the
letters `$1$' and `$2k+3$' into this word in the corresponding symmetric
way we obtain a word representing an element of $S^{\mathrm{HT}}_{2k+3}$.
It is clear that in such a way we obtain all elements of
$S^{\mathrm{HT}}_{2k+3}$ and each element is obtained only once.
If we insert the letter `$1$' before the word and the letter `$2k+3$' after
it, we obtain an element of $S^{\mathrm{HT}}_{2k+3}$ which have the same
number of inversions as the initial element of $S^{\mathrm{HT}}_{2k+1}$.
This gives the contribution equal to $\slantPhi_{\mathrm{HT}}(2k+1; z)$ to
the generating function $\slantPhi_{\mathrm{HT}}(2k+3; z)$. If we insert
the letter `$1$' into the second position and the letter `$2k+3$' into the
next to last position we obtain an element of $S^{\mathrm{HT}}_{2k+3}$
which have the number inversions greater by two as the initial element of
$S^{\mathrm{HT}}_{2k+1}$. This gives a contribution equal to
$\slantPhi_{\mathrm{HT}}(2k+1; z) z^2$. Continuing this procedure we
exhaust all elements of $S^{\mathrm{HT}}_{2k+3}$. Note that when we pass
through the middle of the word the number of inversions increases by three.
Thus, we have
\begin{eqnarray*}
\slantPhi_{\mathrm{HT}}(2k+3; z) &=& \slantPhi_{\mathrm{HT}}(2k+1; z) (1 +
z^2 + \ldots z^{2k} + z^{2k+3} + \ldots + z^{4k+3}) \\
&& \hspace{4em} {} = \slantPhi_{\mathrm{HT}}(2k+1; z)(1 + z^{2k+3})(1 + z^2
+ \ldots + z^{2k}).
\end{eqnarray*}
This equality implies that relation (\ref{e:6}) is valid for $m = k+1$.
Hence, it is valid for any $m \ge 0$. Equality (\ref{e:7}) can be proved in
the same way.
\end{proof}

\subsection{Square-ice model for matrices of even order}

\begin{figure}[ht]
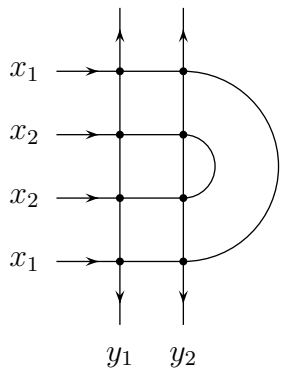

\[
\psset{unit=1em}
\pspicture(-2,-1.4)(7.5,9)
\arrowLine(2,2)(2,0)
\arrowLine(4,2)(4,0)
\arrowLine(2,8)(2,10)
\arrowLine(4,8)(4,10)
\arrowLine(0,2)(2,2)
\arrowLine(0,4)(2,4)
\arrowLine(0,6)(2,6)
\arrowLine(0,8)(2,8)
\psline(2,2)(4,2)
\psline(2,4)(4,4)
\psline(2,6)(4,6)
\psline(2,8)(4,8)
\psline(2,2)(2,8)
\psline(4,2)(4,8)
\psarc(4,5){1}{270}{90}
\psarc(4,5){3}{270}{90}
\psdot(2,2)
\psdot(2,4)
\psdot(2,6)
\psdot(2,8)
\psdot(4,2)
\psdot(4,4)
\psdot(4,6)
\psdot(4,8)
\rput(-1,8){$x_1$}
\rput(-1,6){$x_2$}
\rput(-1,4){$x_2$}
\rput(-1,2){$x_1$}
\rput(2,-1){$y_1$}
\rput(4,-1){$y_2$}
\endpspicture
\]
\caption{Square ice with a half-turn symmetric boundary of even size}
\label{f:hte}
\end{figure}

A method for constructing square-ice models corresponding to
alternating-sign matrices with some symmetry was proposed by Kuperberg
\cite{Kup02}. In this method one actually considers a subset of the
vertices of the state corresponding to the full alternating-sign matrix
which uniquely determines it, and specifies the spectral parameters in an
appropriate convenient way. Kuperberg considered a square-ice model
corresponding to half-turn symmetric alternating-sign matrices of even
order. The structure of the state pattern and the specification of the
spectral parameters for this model can be understood from an example given
in Figure~\ref{f:hte}.\footnote{Actually Kuperberg introduced two variants
of the model differing by specification of the spectral parameters. For our
purposes it suffices to consider only one variant.}
One has the following evident analogue of Lemma \ref{l:2}.

\begin{lemma} \label{l:7}
For a state of square ice with half-turn symmetric boundary of even size
corresponding to the permutation matrix associated with a permutation $s
\in S^\HT_{2m}$, the total number of the vertices of third type and of
forth type is equal to $\inv(s)$ and the total number of the
vertices of fifth type and of sixth type is equal to $m (2m-1) - \inv(s)$.
\end{lemma}

We will denote the partition function of square ice with half-turn
symmetric boundary by $Z_{\mathrm{HT}}(2m, \bm x, \bm
y)$.\footnote{Kuperberg \cite{Kup02} considered only half-turn symmetric
alternating-sign matrices of even order and denoted the partition function
$Z_{\mathrm{HT}}(2m, \bm x, \bm y)$ as $Z_{\mathrm{HT}}(m, \bm x, \bm y)$.}
This function is a Laurent polynomial in the variables $x_i$ and $y_i$.
Introduce the modified partition function
\[
\widetilde Z_{\mathrm{HT}}(2m; \bm x, \bm y) = \left[ \prod_{i=1}^m
x_i^{2m-1} y_i^{2m-1} \right] Z_{\mathrm{HT}}(2m; \bm x, \bm y).
\]

\begin{lemma} \label{l:8}
The modified partition function $\widetilde Z_\HT(2m; \bm x, \bm y)$ has
the following properties.
\begin{itemize}
\item[\rm (a)]
The function $\widetilde Z_\HT(2m; \bm x, \bm y)$ is symmetric separately
in the variables $x_1, \ldots, x_m$ and in the variables $y_1, \ldots,
y_m$.
\item[\rm (b)]
The function $\widetilde Z_\HT(2m; \bm x, \bm y)$ is a homogeneous
polynomial in the variables $x_i$ and $y_i$ of total degree $2 m (2m-1)$.
For each fixed $i = 1, \ldots, m$ it is a polynomial in $x_i^2$ of degree
$2m-1$ and a polynomial in $y_i^2$ of degree $2m-1$. 
\item[\rm (c)]
If $y_m = a x_m$, then
\begin{eqnarray*}
\lefteqn{\widetilde Z_\HT(2m; \bm x, \bm y) = \sigma^2(a^2) \, x_m y_m} \\
&& \hspace{2em} \prod_{i=1}^{m-1} \left[(a y_i^2 - \bar a x_m^2)^2 (a y_m^2
- \bar a x_i^2)^2 \right] \widetilde Z_\HT(2m-2; \bm x \smallsetminus
x_m, \bm y \smallsetminus y_m).
\end{eqnarray*}
\end{itemize}
\end{lemma}

\begin{proof} 
The stated properties of the modified partition function $\widetilde
Z_\HT(2m; \bm x, \bm y)$ follow from the corresponding properties of the
partition function $Z_\HT(2m; \bm x, \bm y)$, see \cite{Kup02}.
\end{proof}

Singling out from $\widetilde Z_\HT(2m; \bm x, \bm y)$ the term of maximal
degree in the variables $x_i$, we write
\[
\widetilde Z_\HT(2m; \bm x, \bm y) = \left[ \prod_{i=1}^m x_i^{2(2m-1)}
\right] C_\HT(2m) + \ldots .
\]

\begin{lemma} \label{l:9}
The coefficient $C_\HT(2m)$ is given by the formula
\begin{equation}
C_\HT(2m) = \prod_{i=1}^{2m} \sigma(a^i). \label{e:8}
\end{equation}
\end{lemma}

\begin{proof}
It is clear that only the states corresponding to permutation matrices
contribute to the term of maximal degree in the variables $x_i$.
Using the same reasonings as in the proof of Lemma~\ref{l:4} and taking
into account Lemma~\ref{l:7}, we see that the contribution of a state
corresponding to a permutation $s$ is
\begin{eqnarray*}
\lefteqn{\left[ \prod_{i=1}^m x_i^{2(2m-1)} \right] \sigma^m(a^2)
a^{\inv(s)} (-\bar a) ^{m(2m-1) - \inv(s)}} \\
&=& \left[ \prod_{i=1}^m x_i^{2(2m-1)} \right] \sigma^m(a^2)
(-a)^{-m(2m-1)} (-a^2)^{\inv(s)}.
\end{eqnarray*}
Thus, we have
\[
C_\HT(2m) = (-1)^m \sigma^m(a^2) a^{-m(2m-1)} \sum_{s \in S^\HT_{2m}}
(-a^2)^{\inv(s)}.
\]
Using relation (\ref{e:7}), we come to equality (\ref{e:8}).
\end{proof}

Singling out the terms of $\widetilde Z_\HT(2m; \bm x, \bm y)$ which have
the maximal total degree in all the variables $x_i$ except the variable
$x_m$, and do not contain all the variables $y_i$ except the variable
$y_m$, we write
\[
\widetilde Z_\HT(2 m; \bm x, \bm y) = \left[ \prod_{i=1}^{m-1}
x_i^{2(2m-1)} \right] S_\HT(2m; x_m, y_m) + \ldots .
\]
To find the polynomial $S_\HT(2m, x_m, y_m)$ we use the fact
proved by Kuperberg \cite{Kup02} that the partition function
$Z_{\mathrm{HT}}(2m, \bm x, \bm y)$ is the product of two Laurent
polynomials,
\[
Z_{\mathrm{HT}}(2m; \bm x,\bm y) = Z(m; \bm  x, \bm  y)
Z_{\mathrm{HT}}^{(2)}(2m; \bm x, \bm y).
\]
Hence, one can write
\begin{equation}
\widetilde Z_{\mathrm{HT}}(2m; \bm x,\bm y) = \widetilde Z(m; \bm  x, \bm
y) \widetilde Z_{\mathrm{HT}}^{(2)}(2m; \bm x, \bm y). \label{e:9}
\end{equation}
\begin{lemma}
The function $\widetilde Z_\HT^{(2)}(2m; \bm x, \bm y)$ is a
homogeneous polynomial in the variables $x_i$ and $y_i$ of total degree $2
m^2$. For each fixed $i = 1, \ldots, m$ it is a polynomial in $x_i^2$ of
degree $m$ and a polynomial in $y_i^2$ of degree $m$. If $y_m = a x_m$,
then
\begin{eqnarray}
\lefteqn{\widetilde Z_\HT^{(2)}(2m; \bm x, \bm y) = \sigma(a^2) \, x_m y_m}
\nonumber \\
&& \hspace{2em} {} \times \prod_{i=1}^{m-1} \left[ (a y_i^2 - \bar a x_m^2)
(a y_m^2 - \bar a x_i^2) \right] \widetilde Z_\HT^{(2)}(2(m-1); \bm x
\smallsetminus x_m, \bm y \smallsetminus y_m). \label{e:10}
\end{eqnarray}
\end{lemma}

\begin{proof} One can prove the lemma using relation (\ref{e:9}) and Lemmas
\ref{l:8} and \ref{l:3}.
\end{proof}

\begin{lemma}
The polynomial $S_\HT(2m; x_m, y_m)$ has the form
\begin{equation}
S_\HT(2m; x_m, y_m) = S(m, x_m, y_m) S_\HT^{(2)}(2m, x_m, y_m),
\label{e:11}
\end{equation}
where $S(m, x_m, y_m)$ is given by relation (\ref{e:5}) and
\begin{equation}
S_\HT^{(2)}(2m; x_m, y_m) = \left[ \prod_{i=1}^{m-1} \sigma(a^{2i-1})
\right] \left( \sigma(a^{2m-1}) x_m^{2m} + \sigma(a^{2m-3}) x_m^{2(m-1)}
y_m^2 \right). \label{e:12}
\end{equation}
\end{lemma}

\begin{proof}
Consider a state of square ice with half-turn symmetric boundary which
gives a nontrivial contribution to the polynomial $S_\HT(2m; x_m, y_m)$. It
is clear that for such a state each line with the spectral parameter $y_j$,
$j = 1, \ldots, m-1$ may have only one vertex of first or second type, and
the same is true for each line with the spectral parameter $x_i$, $i = 1,
\ldots, m-1$. For the line with the spectral parameter $x_m$ we have two
possibilities. Either it has only one vertex of first or second type, or it
has two vertices of first type and one of second type. In the latter case
the line with the spectral parameter $y_m$ also has two vertices of first
type and one vertex of second type. Analysing all the possibilities one
concludes that the polynomial
$S_\HT(2m; x_m, y_m)$ has the form
\[
S_\HT(2m; x_m, y_m) = C_\HT(2m) x_m^{2(2m-1)} + D_\HT(2m) x_m^{2(2m-2)}
y_m^2 + E_\HT(2m) x_m^{2(2m-3)} y_m^4.
\]
It follows from equality (\ref{e:9}) that the polynomial $S_\HT(2m, x_m,
y_m)$ can be represented in form (\ref{e:11}) for some polynomial
$S_\HT^{(2)}(2m; x_m, y_m)$. Hence, having in mind (\ref{e:5}), one can see
that
\[
S_\HT^{(2)}(2m; x_m, y_m) = C_\HT^{(2)}(2m) x_m^{2 m} + D_\HT^{(2)}(2m)
x_m^{2(m-1)} y_m^2.
\]
Using the evident equality
\[
C_\HT(2m) = C(m) C_\HT^{(2)}(2m),
\]
one obtains the relation
\begin{equation}
C_\HT^{(2)}(2m) = \prod_{i=1}^m \sigma(a^{2i-1}). \label{e:13}
\end{equation}
Recursive relation (\ref{e:10}) gives
\[
C_\HT^{(2)}(2m) + a^2 D_\HT^{(2)}(2m) = \sigma(a^2) a^{-2m+3}
C_\HT^{(2)}(2(m-1)).
\]
Taking into account equality (\ref{e:8}), one comes to the relation
\[
D_\HT^{(2)}(2m) = - \left[ \prod_{i=1}^{m-1} \sigma(a^{2i-1}) \right]
\sigma(a^{2m-3})
\]
which implies equality (\ref{e:12}).
\end{proof}

\subsection{Square-ice model for matrices of odd order}

\begin{figure}[ht]
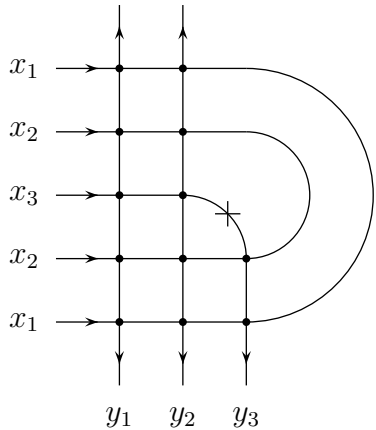

\[
\psset{unit=1em}
\pspicture(-2,-1.4)(10,12)
\arrowLine(2,2)(2,0)
\arrowLine(4,2)(4,0)
\arrowLine(2,10)(2,12)
\arrowLine(4,10)(4,12)
\arrowLine(0,2)(2,2)
\arrowLine(0,4)(2,4)
\arrowLine(0,8)(2,8)
\arrowLine(0,10)(2,10)
\psline(2,2)(6,2)
\psline(2,4)(6,4)
\psline(2,8)(6,8)
\psline(2,10)(6,10)
\psline(2,2)(2,10)
\psline(4,2)(4,10)
\psarc(6,6){2}{270}{90}
\psarc(6,6){4}{270}{90}
\psdot(2,2)
\psdot(2,4)
\psdot(2,8)
\psdot(2,10)
\psdot(4,2)
\psdot(4,4)
\psdot(4,8)
\psdot(4,10)
\psdot(2,6)
\psdot(4,6)
\psdot(6,4)
\psdot(6,2)
\arrowLine(0,6)(2,6)
\psline(2,6)(4,6)
\psarc(4,4){2}{0}{90}
\psline(5.014,5.414)(5.814,5.414)
\psline(5.414,5.014)(5.414,5.814)
\psline(6,4)(6,2)
\arrowLine(6,2)(6,0)
\rput(-1,10){$x_1$}
\rput(-1,8){$x_2$}
\rput(-1,6){$x_3$}
\rput(-1,4){$x_2$}
\rput(-1,2){$x_1$}
\rput(2,-1){$y_1$}
\rput(4,-1){$y_2$}
\rput(6,-1){$y_3$}
\endpspicture
\]
\caption{Square ice with a half-turn symmetric boundary of odd size}
\label{f:hto}
\end{figure}

Proceed now to the case of half-turn symmetric alternating-sign matrices of
odd order. It is useful to have in mind that the central matrix element of
such a matrix is either 1 or $-1$. The structure of the state pattern and
the specification of the spectral parameters in this case can be understood
from an example given in Figure~\ref{f:hto}. The cross means the change of
the spectral parameter associated with the line. Here the direction of the
arrow should be preserved. The following lemma is evident.

\begin{lemma} \label{l:12}
For a state of square ice with half-turn symmetric boundary of odd
size corresponding to the permutation matrix associated with a permutation
$s \in S^\HT_{2m+1}$, the total number of the vertices of third type
and of forth type is equal to $\inv(s)$ and the total number of the
vertices of fifth type and of sixth type is equal to $m(2m+1) - \inv(s)$.
\end{lemma}

The partition function $Z_\HT(2m+1; \bm x, \bm y)$ is a Laurent polynomial
in the variables $x_i$ and $y_i$. It is convenient to introduce the
modified partition function
\[
\widetilde Z_{\mathrm{HT}}(2m+1; \bm x, \bm y) = \left[ \prod_{i=1}^m
x_i^{2m} y_i^{2m} \right] x_{m+1}^m y_{m+1}^m Z_{\mathrm{HT}}(2m+1; \bm x,
\bm y)
\]
which is a polynomial in the spectral parameters.

\begin{lemma} \label{l:13}
The partition function $Z_\HT(2m + 1; \bm x, \bm y)$ is symmetric
separately in the variables $x_1, \ldots, x_m$ and in the variables
$y_1, \ldots, y_m$. If $y_1 = a x_1$, then
\begin{eqnarray}
\lefteqn{Z_\HT(2m + 1; \bm x, \bm y) = \sigma^2(a^2) \sigma(a \bar x_1
y_{m+1}) \sigma(a \bar x_{m+1} y_1)} \nonumber \\
&& \hspace{6em} {} \times \prod_{i=2}^m \left[ \sigma^2(a \bar
x_1 y_i) \sigma^2 (a \bar x_i y_1) \right] Z_\HT(2m-1; \bm x
\smallsetminus x_1, \bm y \smallsetminus y_1). \label{e:14}
\end{eqnarray}
\end{lemma}

\begin{proof}
To prove the symmetricity of the partition sum $Z_\HT(2m + 1; \bm x, \bm
y)$ in the variables $x_1$ and $x_2$ one multiplies it by $\sigma(az)$. The
resulting expression may be considered as corresponding to the graph at the
left-hand side of the equality given in Figure \ref{f:sym}.
\begin{figure}[ht]
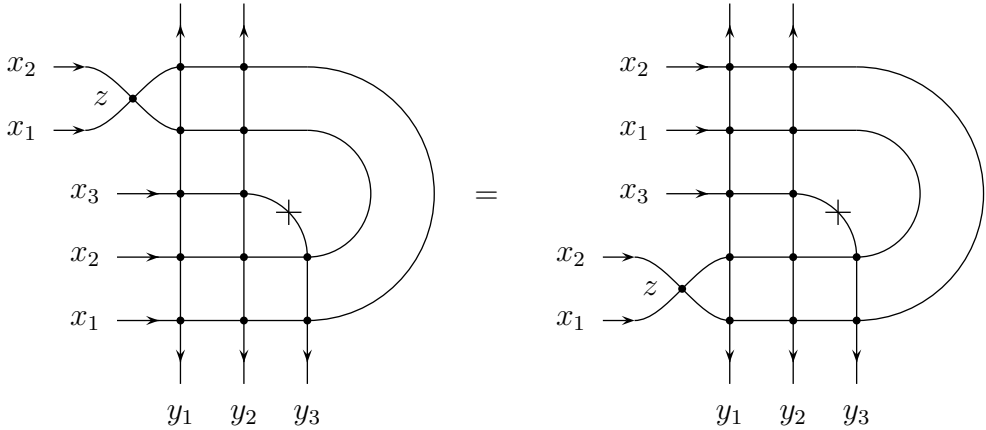

\[
\psset{unit=1em}
\pspicture[.53](-4,-1.4)(10,12)
\arrowLine(2,2)(2,0)
\arrowLine(4,2)(4,0)
\arrowLine(2,10)(2,12)
\arrowLine(4,10)(4,12)
\arrowLine(0,2)(2,2)
\arrowLine(0,4)(2,4)
\psline[arrows=>](-1,8)(-2,8)
\psline[arrows=>](-1,10)(-2,10)
\psbezier(-1,10)(-0,10)(1,8)(2,8)
\psbezier(-1,8)(-0,8)(1,10)(2,10)
\psline(2,2)(6,2)
\psline(2,4)(6,4)
\psline(2,8)(6,8)
\psline(2,10)(6,10)
\psline(2,2)(2,10)
\psline(4,2)(4,10)
\psarc(6,6){2}{270}{90}
\psarc(6,6){4}{270}{90}
\psdot(.5,9)
\psdot(2,2)
\psdot(2,4)
\psdot(2,8)
\psdot(2,10)
\psdot(4,2)
\psdot(4,4)
\psdot(4,8)
\psdot(4,10)
\psdot(2,6)
\psdot(4,6)
\psdot(6,4)
\psdot(6,2)
\arrowLine(0,6)(2,6)
\psline(2,6)(4,6)
\psarc(4,4){2}{0}{90}
\psline(5.014,5.414)(5.814,5.414)
\psline(5.414,5.014)(5.414,5.814)
\psline(6,4)(6,2)
\arrowLine(6,2)(6,0)
\rput(-3,10){$x_2$}
\rput(-3,8){$x_1$}
\rput(-1,6){$x_3$}
\rput(-1,4){$x_2$}
\rput(-1,2){$x_1$}
\rput(2,-1){$y_1$}
\rput(4,-1){$y_2$}
\rput(6,-1){$y_3$}
\rput(-.5,9){$z$}
\endpspicture 
\quad = \quad
\pspicture[.53](-4,-1.4)(10,12)
\arrowLine(2,2)(2,0)
\arrowLine(4,2)(4,0)
\arrowLine(2,10)(2,12)
\arrowLine(4,10)(4,12)
\psline[arrows=>](-1,2)(-2,2)
\psline[arrows=>](-1,4)(-2,4)
\psbezier(-1,4)(-0,4)(1,2)(2,2)
\psbezier(-1,2)(-0,2)(1,4)(2,4)
\arrowLine(0,8)(2,8)
\arrowLine(0,10)(2,10)
\psline(2,2)(6,2)
\psline(2,4)(6,4)
\psline(2,8)(6,8)
\psline(2,10)(6,10)
\psline(2,2)(2,10)
\psline(4,2)(4,10)
\psarc(6,6){2}{270}{90}
\psarc(6,6){4}{270}{90}
\psdot(.5,3)
\psdot(2,2)
\psdot(2,4)
\psdot(2,8)
\psdot(2,10)
\psdot(4,2)
\psdot(4,4)
\psdot(4,8)
\psdot(4,10)
\psdot(2,6)
\psdot(4,6)
\psdot(6,4)
\psdot(6,2)
\arrowLine(0,6)(2,6)
\psline(2,6)(4,6)
\psarc(4,4){2}{0}{90}
\psline(5.014,5.414)(5.814,5.414)
\psline(5.414,5.014)(5.414,5.814)
\psline(6,4)(6,2)
\arrowLine(6,2)(6,0)
\rput(-1,10){$x_2$}
\rput(-1,8){$x_1$}
\rput(-1,6){$x_3$}
\rput(-3,4){$x_2$}
\rput(-3,2){$x_1$}
\rput(2,-1){$y_1$}
\rput(4,-1){$y_2$}
\rput(6,-1){$y_3$}
\rput(-.5,3){$z$}
\endpspicture
\]
\caption{The proof of the symmetricity of $Z_\HT(3; \bm x, \bm y)$ in
$x_1$ and $x_2$}
\label{f:sym}
\end{figure}
If $z = a \bar x_1 x_2$, then using the Yang--Baxter equation, see
Figure \ref{f:yb}, one can move the crossing to the position
given in the graph at the left-hand side of this equality. This proves the
symmetricity in $x_1$ and $x_2$. The other variables are treated in the
same way.

To prove recursive relation (\ref{e:14}) we note that if $y_1 = a x_1$,
then only the states with a vertex of first type in the top-left corner
give nonzero contribution to the partition function. Here the tetravalent
vertices at the boundary of the graph become fixed, see Figure \ref{f:rr}.
\begin{figure}[ht]
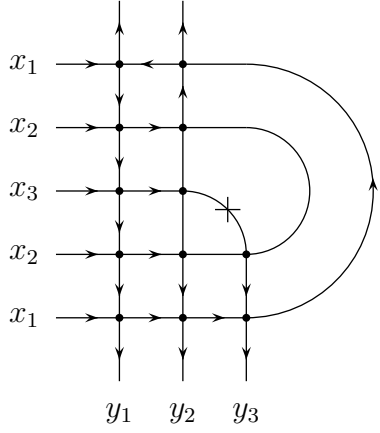

\[
\psset{unit=1em}
\pspicture(-2,-1.4)(10,12)
\arrowLine(2,2)(2,0)
\arrowLine(4,2)(4,0)
\arrowLine(2,10)(2,12)
\arrowLine(4,10)(4,12)
\arrowLine(0,2)(2,2)
\arrowLine(0,4)(2,4)
\arrowLine(0,8)(2,8)
\arrowLine(0,10)(2,10)
\arrowLine(2,2)(4,2)
\arrowLine(4,2)(6,2)
\arrowLine(2,4)(4,4)
\psline(4,4)(6,4)
\arrowLine(2,8)(4,8)
\psline(4,8)(6,8)
\arrowLine(4,10)(2,10)
\psline(4,10)(6,10)
\arrowLine(2,10)(2,8)
\arrowLine(2,8)(2,6)
\arrowLine(2,6)(2,4)
\arrowLine(2,4)(2,2)
\arrowLine(4,4)(4,2)
\arrowLine(4,8)(4,10)
\psline(4,4)(4,8)
\psarc(6,6){2}{270}{90}
\psarc[arrows=<](6,6){4}{0}{90}
\psarc(6,6){4}{270}{5}
\psdot(2,2)
\psdot(2,4)
\psdot(2,8)
\psdot(2,10)
\psdot(4,2)
\psdot(4,4)
\psdot(4,8)
\psdot(4,10)
\psdot(2,6)
\psdot(4,6)
\psdot(6,4)
\psdot(6,2)
\arrowLine(0,6)(2,6)
\arrowLine(2,6)(4,6)
\psarc(4,4){2}{0}{90}
\psline(5.014,5.414)(5.814,5.414)
\psline(5.414,5.014)(5.414,5.814)
\arrowLine(6,4)(6,2)
\arrowLine(6,2)(6,0)
\rput(-1,10){$x_1$}
\rput(-1,8){$x_2$}
\rput(-1,6){$x_3$}
\rput(-1,4){$x_2$}
\rput(-1,2){$x_1$}
\rput(2,-1){$y_1$}
\rput(4,-1){$y_2$}
\rput(6,-1){$y_3$}
\endpspicture
\]
\caption{The proof of recursive relation (\ref{e:14})}
\label{f:rr}
\end{figure}
They give all but last factors in the right-hand side of (\ref{e:14}), and
the remaining vertices give the last factor.
\end{proof}

\begin{lemma} \label{l:14}
The modified partition function $\widetilde Z_\HT(2m+1; \bm x, \bm y)$ has
the following properties.
\begin{itemize}
\item[\rm (a)]
The function $\widetilde Z_\HT(2m + 1; \bm x, \bm y)$ is symmetric
separately in the variables $x_1, \ldots, x_m$ and in the variables $y_1,
\ldots, y_m$.
\item[\rm (b)]
The function $\widetilde Z_\HT(2m+1; \bm x, \bm y)$ is a homogeneous
polynomial in the variables $x_i$ and $y_i$ of total degree $2 m (2m+1)$.
For each fixed $i = 1, \ldots, m$ it is a polynomial in $x_i^2$ of degree
$2m$ and a polynomial in $y_i^2$ of degree $2m$; it is a polynomial of
degree $2m$ in $x_{m+1}$ and a polynomial of degree $2m$ in $y_{m+1}$. 
\item[\rm (c)]
If $y_m = a x_m$, then
\begin{eqnarray}
\lefteqn{\widetilde Z_\HT(2m + 1; \bm x, \bm y) = \sigma^2(a^2) (a
y_{m+1}^2 - \bar a x_m^2) (a y_m^2 - \bar a x_{m+1}^2) x_{m} y_m} \nonumber
\\
&& \hspace{2em} {} \times \prod_{i=1}^{m-1} \left[ (a y_i^2 - \bar
a x_m^2)^2 (a y_m^2 - \bar a x_i^2)^2 \right] \widetilde Z_\HT(2m-1; \bm x
\smallsetminus x_m, \bm y \smallsetminus y_m). \hspace{2em} \label{e:15}
\end{eqnarray}
\end{itemize}
\end{lemma}

\begin{proof}
The first two statements of the lemma are evident. Recursive relation
(\ref{e:15}) follows from the symmetricity of $Z_\HT(2m + 1; \bm x, \bm y)$
and recursive relation (\ref{e:14}).
\end{proof}

Singling out from $\widetilde Z_\HT(2m+1; \bm x, \bm y)$ the term of
maximal degree in the variables $x_i$, we write
\[
\widetilde Z_\HT(2m+1; \bm x, \bm y) = \left[ \prod_{i=1}^m x_i^{4m}
\right] x_{m+1}^{2m} C_\HT(2m+1) + \ldots .
\]

\begin{lemma}
The coefficient $C_\HT(2m+1)$ is given by the formula
\begin{equation}
C_\HT(2m+1) = \prod_{i=2}^{2m+1} \sigma(a^i). \label{e:16}
\end{equation}
\end{lemma}

\begin{proof}
One can prove the lemma in the same way as Lemma \ref{l:9} using Lemma
\ref{l:12} and equality~(\ref{e:6}).
\end{proof}

Singling out the terms of $\widetilde Z_\HT(2m+1; \bm x, \bm y)$ which have
the maximal total degree in all spectral parameters except $x_{m+1}$ and
$y_{m+1}$, we write
\[
\widetilde Z_\HT(2m+1; \bm x, \bm y) = \left[ \prod_{i=1}^m
x_i^{4m} \right] S_\HT(2m+1; x_{m+1}, y_{m+1}) + \ldots.
\]
The properties of the modified partition functions $\widetilde Z(n; \bm x,
\bm y)$ and $\widetilde Z_\HT(2m, \bm x, \bm y)$ described by Lemmas
\ref{l:3} and \ref{l:8}, respectively, determine them uniquely by Lagrange
interpolation. It is not the case for the modified partition function
$\widetilde Z_\HT(2m+1, \bm x, \bm y)$. Actually, what is missed here is
the polynomial $S_\HT(2m+1, x_{m+1}, y_{m+1})$. To prove this fact we need
the following lemma.

\begin{lemma}
The partition function $Z_\HT(2m+1; \bm x, \bm y)$ is invariant under the
replacement $x_i \to \bar x_i$, $y_i \to \bar y_i$. For the modified
partition function on has
\begin{equation}
\widetilde Z_{\mathrm{HT}}(2m+1; \bar{\bm x}, \bar{\bm y}) = \left[
\prod_{i=1}^m x_i^{-4m} y_i^{-4m} \right] x_{m+1}^{-2m} \, y_{m+1}^{-2m} \,
\widetilde Z_{\mathrm{HT}}(2m+1; \bm x, \bm y), \label{e:17}
\end{equation}
where $\bar{\bm x} = (\bar x_1, \ldots, \bar x_{m+1})$ and $\bar{\bm y} =
(\bar y_1, \ldots, \bar y_{m+1})$.
\end{lemma}

\begin{proof}
Consider a state of square ice with a half-turn symmetric boundary.
Reflect the corresponding graph through a horizontal line, and then rotate
the half-line with the spectral parameter $y_{m+1}$ by $180^\circ$. It is
clear that the weight of the new state is obtained from the weight of the
old one by the substitution $x_i \to \bar x_i$, $y_i \to \bar y_i$. This
fact implies the invariance of $Z_\HT(2m+1; \bm x, \bm y)$ under this
substitution. Equality (\ref{e:17}) follows immediately from the invariance
of $Z_\HT(2m+1; \bm x, \bm y)$.
\end{proof}

\begin{lemma} \label{l:17}
If two functions $\widetilde Z_\HT(2m+1; \bm x, \bm y)$ and $\widetilde
Z'_\HT(2m+1; \bm x, \bm y)$ have the properties described in Lemma
\ref{l:14}, and the corresponding polynomials $S_\HT(2m+1; x_{m+1},
y_{m+1})$ and $S'_\HT(2m+1; x_{m+1}, y_{m+1})$ coincide, then $\widetilde
Z_\HT(2m+1; \bm x, \bm y) = \widetilde Z'_\HT(2m+1; \bm x, \bm y)$.
\end{lemma}

\begin{proof}
Using statements (a) and (b) of Lemma \ref{l:14}, one can see that the
difference of $\widetilde Z_\HT(2m+1; \bm x, \bm y)$ and $\widetilde
Z'_\HT(2m+1; \bm x, \bm y)$ is zero when $y_j = a x_i$. Since $\widetilde
Z_\HT(2m+1; \bm x, \bm y)$ and $\widetilde Z'_\HT(2m+1; \bm x, \bm y)$ are
polynomial in $y_i^2$, this difference is also zero when $y_j = - a x_i$.
Using (\ref{e:17}), one can put recursive relation (\ref{e:15}) into the
form valid for $y_1 = \bar a x_1$. This form of the recursive relation
(\ref{e:15}) and statement (a) of Lemma \ref{l:14} implies that the
difference of $\widetilde Z_\HT(2m+1; \bm x, \bm y)$ and $\widetilde
Z'_\HT(2m+1; \bm x, \bm y)$ is zero when $y_j = \bar a x_i$ and $y_j = \bar
a x_i$. Using these facts one easily obtains the equality
\begin{eqnarray*}
\lefteqn{\widetilde Z_\HT(2m+1; \bm x, \bm y) - \widetilde Z'_\HT(2m+1; \bm
x, \bm y) = \prod_{i,j = 1}^m \left[ (y_i^2 - a^2 x_j) (y_i^2 - \bar a^2
x_j) \right]} \\
&& \hspace{8em} {} \times (S_\HT(2m+1; x_{m+1}, y_{m+1}) - S'_\HT(2m+1;
x_{m+1}, y_{m+1})),
\end{eqnarray*}
which makes the statement of the lemma evident.
\end{proof}

\subsection{Additional recursive relation}

To find the polynomial $S_\HT(2m+1; x_{m+1}, y_{m+1})$ we use an additional
recursive relation satisfied by the partition function $Z_\HT(2m+1; \bm x,
\bm y)$.

\begin{lemma}
If $y_{m+1} = a x_{m+1}$, then
\begin{equation}
Z_{\mathrm{HT}}(2m + 1; \bm x, \bm y) = \prod_{i=1}^m \left[ \sigma(a
\bar x_i y_{m+1}) \sigma (a \bar x_{m+1} y_i) \right]
Z_{\mathrm{HT}}(2m; \bm x \smallsetminus x_{m+1}, \bm y \smallsetminus
y_{m+1}). \label{e:18}
\end{equation}
\end{lemma}

\begin{proof}
Consider the spectral parameters of the vertices belonging to the central
`triangle' of the graph describing a state of square ice with a half-turn
symmetric boundary, see example in Figure \ref{f:ct}.
\begin{figure}[ht]
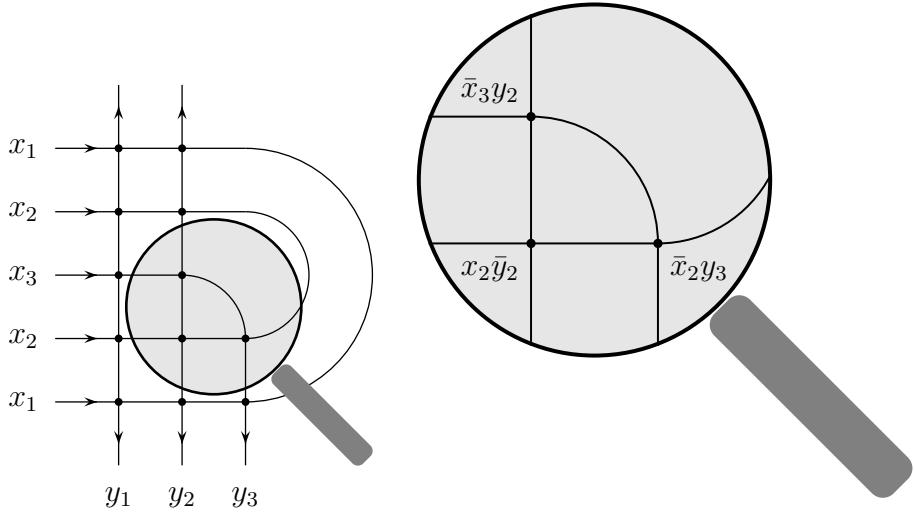

\[
\psset{unit=1em}
\newgray{mygray}{.9}
\pspicture(-2,-1.4)(26,15)
\pscircle*[linecolor=mygray](5,5){2.8}
\pscircle[linewidth=1pt](5,5){2.8}
\pscircle*[linecolor=mygray](17,9){5.6}
\pscircle[linewidth=1.5pt](17,9){5.6}
\arrowLine(2,2)(2,0)
\arrowLine(4,2)(4,0)
\arrowLine(2,10)(2,12)
\arrowLine(4,10)(4,12)
\arrowLine(0,2)(2,2)
\arrowLine(0,4)(2,4)
\arrowLine(0,8)(2,8)
\arrowLine(0,10)(2,10)
\psline(2,2)(6,2)
\psline(2,4)(6,4)
\psline(2,8)(6,8)
\psline(2,10)(6,10)
\psline(2,2)(2,10)
\psline(4,2)(4,10)
\psarc(6,6){2}{270}{90}
\psarc(6,6){4}{270}{90}
\psdot(2,2)
\psdot(2,4)
\psdot(2,8)
\psdot(2,10)
\psdot(4,2)
\psdot(4,4)
\psdot(4,8)
\psdot(4,10)
\psdot(2,6)
\psdot(4,6)
\psdot(6,4)
\psdot(6,2)
\arrowLine(0,6)(2,6)
\psline(2,6)(4,6)
\psarc(4,4){2}{0}{90}
\psline(6,4)(6,2)
\arrowLine(6,2)(6,0)
\rput(-1,10){$x_1$}
\rput(-1,8){$x_2$}
\rput(-1,6){$x_3$}
\rput(-1,4){$x_2$}
\rput(-1,2){$x_1$}
\rput(2,-1){$y_1$}
\rput(4,-1){$y_2$}
\rput(6,-1){$y_3$}
\psset{linewidth=.75pt}
\psline(11.77,7)(19,7)
\psline(11.77,11)(15,11)
\psline(15,3.77)(15,14.23)
\psline(19,3.77)(19,7)
\psarc(15,7){4}{0}{90}
\psarc(19,11){4}{-90}{-27}
\psdot(15,7)
\psdot(15,11)
\psdot(19,7)
\rput(13.7,6.2){$x_2 \bar y_2$}
\rput(13.7,11.8){$\bar x_3 y_2$}
\rput(20.3,6.2){$\bar x_2 y_3$}
\rput{225}(7,3){%
\psframe*[framearc=.6,fillcolor=gray,linecolor=gray](-.4,0)(.4,4)}
\rput{225}(21,5){%
\psframe*[framearc=.6,fillcolor=gray,linecolor=gray](-.8,0)(.8,8)}
\endpspicture
\]
\caption{The central `triangle' under a magnifying glass}
\label{f:ct}
\end{figure}
It is clear that if $y_{m+1} = a x_{m+1}$ then we can use the Yang--Baxter
equation, given in Figure \ref{f:yb}, and move the line with spectral
parameters $x_{m+1}$ and $y_{m+1}$ to the boundary. An example of the
process is given in Figure \ref{f:arr}.
\begin{figure}[ht]
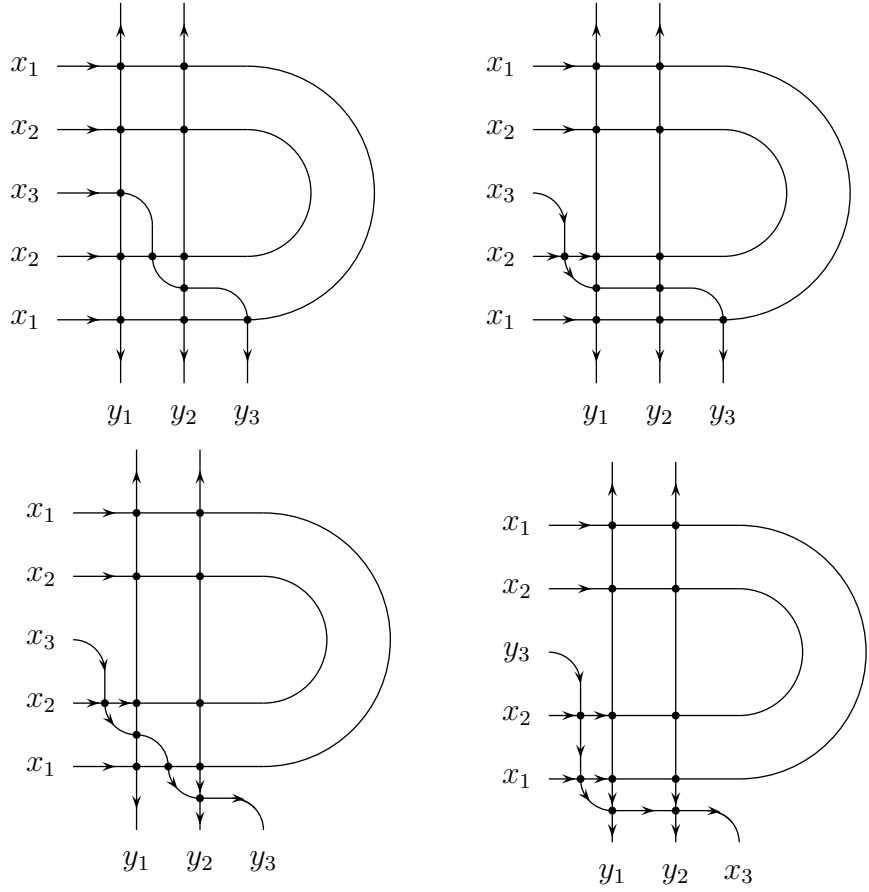

\[
\psset{unit=1em}
\pspicture(-1,-1)(10,12)
\arrowLine(2,2)(2,0)
\arrowLine(4,2)(4,0)
\arrowLine(2,10)(2,12)
\arrowLine(4,10)(4,12)
\arrowLine(0,2)(2,2)
\arrowLine(0,4)(2,4)
\arrowLine(0,8)(2,8)
\arrowLine(0,10)(2,10)
\psline(2,2)(6,2)
\psline(2,4)(6,4)
\psline(2,8)(6,8)
\psline(2,10)(6,10)
\psline(2,2)(2,10)
\psline(4,2)(4,10)
\psarc(6,6){2}{270}{90}
\psarc(6,6){4}{270}{90}
\psdot(2,2)
\psdot(2,4)
\psdot(2,8)
\psdot(2,10)
\psdot(4,2)
\psdot(4,4)
\psdot(4,8)
\psdot(4,10)
\psdot(2,6)
\psdot(3,4)
\psdot(4,3)
\psdot(6,2)
\arrowLine(0,6)(2,6)
\psarc(2,5){1}{0}{90}
\psline(3,5)(3,4)
\psarc(4,4){1}{180}{270}
\psline(4,3)(5,3)
\psarc(5,2){1}{0}{90}
\arrowLine(6,2)(6,0)
\rput(-1,10){$x_1$}
\rput(-1,8){$x_2$}
\rput(-1,6){$x_3$}
\rput(-1,4){$x_2$}
\rput(-1,2){$x_1$}
\rput(2,-1){$y_1$}
\rput(4,-1){$y_2$}
\rput(6,-1){$y_3$}
\endpspicture
\qquad \qquad
\pspicture(-1,-1)(10,12)
\arrowLine(2,2)(2,0)
\arrowLine(4,2)(4,0)
\arrowLine(2,10)(2,12)
\arrowLine(4,10)(4,12)
\arrowLine(0,2)(2,2)
\arrowLine(0,8)(2,8)
\arrowLine(0,10)(2,10)
\psline(2,2)(6,2)
\psline(2,4)(6,4)
\psline(2,8)(6,8)
\psline(2,10)(6,10)
\psline(2,2)(2,10)
\psline(4,2)(4,10)
\psarc(6,6){2}{270}{90}
\psarc(6,6){4}{270}{90}
\psdot(2,2)
\psdot(2,4)
\psdot(2,8)
\psdot(2,10)
\psdot(4,2)
\psdot(4,4)
\psdot(4,8)
\psdot(4,10)
\psdot(1,4)
\psdot(2,3)
\psdot(4,3)
\psdot(6,2)
\psarc{<-}(0,5){1}{0}{90}
\psline(1,5.2)(1,4)
\psarc{->}(2,4){1}{180}{225}
\psarc(2,4){1}{220}{270}
\arrowLine(0,4)(1,4)
\arrowLine(1,4)(2,4)
\psline(2,3)(5,3)
\psarc(5,2){1}{0}{90}
\arrowLine(6,2)(6,0)
\rput(-1,10){$x_1$}
\rput(-1,8){$x_2$}
\rput(-1,6){$x_3$}
\rput(-1,4){$x_2$}
\rput(-1,2){$x_1$}
\rput(2,-1){$y_1$}
\rput(4,-1){$y_2$}
\rput(6,-1){$y_3$}
\endpspicture
\]
\[
\psset{unit=1em}
\pspicture(-2,-1.4)(10,12)
\arrowLine(2,2)(2,0)
\arrowLine(4,2)(4,1)
\arrowLine(4,1)(4,0)
\arrowLine(2,10)(2,12)
\arrowLine(4,10)(4,12)
\arrowLine(0,2)(2,2)
\arrowLine(0,4)(1,4)
\arrowLine(1,4)(2,4)
\arrowLine(0,8)(2,8)
\arrowLine(0,10)(2,10)
\psline(2,2)(6,2)
\psline(2,4)(6,4)
\psline(2,8)(6,8)
\psline(2,10)(6,10)
\psline(2,2)(2,10)
\psline(4,2)(4,10)
\psarc(6,6){2}{270}{90}
\psarc(6,6){4}{270}{90}
\psdot(2,2)
\psdot(2,4)
\psdot(2,8)
\psdot(2,10)
\psdot(4,2)
\psdot(4,4)
\psdot(4,8)
\psdot(4,10)
\psdot(1,4)
\psdot(2,3)
\psdot(3,2)
\psdot(4,1)
\psarc{<-}(0,5){1}{0}{90}
\psline(1,5.2)(1,4)
\psarc{->}(2,4){1}{180}{225}
\psarc(2,4){1}{220}{270}
\psarc(2,2){1}{0}{90}
\psarc{->}(4,2){1}{180}{225}
\psarc(4,2){1}{220}{270}
\psline(4,1)(5.2,1)
\psarc{-<}(5,0){1}{0}{90}
\psarc(5,0){1}{40}{90}
\rput(-1,10){$x_1$}
\rput(-1,8){$x_2$}
\rput(-1,6){$x_3$}
\rput(-1,4){$x_2$}
\rput(-1,2){$x_1$}
\rput(2,-1){$y_1$}
\rput(4,-1){$y_2$}
\rput(6,-1){$y_3$}
\endpspicture
\qquad \qquad
\pspicture(-1,-1)(10,12)
\arrowLine(2,2)(2,1)
\arrowLine(2,1)(2,0)
\arrowLine(4,2)(4,1)
\arrowLine(4,1)(4,0)
\arrowLine(2,10)(2,12)
\arrowLine(4,10)(4,12)
\arrowLine(0,2)(1,2)
\arrowLine(1,2)(2,2)
\arrowLine(0,4)(1,4)
\arrowLine(1,4)(2,4)
\arrowLine(0,8)(2,8)
\arrowLine(0,10)(2,10)
\psline(2,2)(6,2)
\psline(2,4)(6,4)
\psline(2,8)(6,8)
\psline(2,10)(6,10)
\psline(2,2)(2,10)
\psline(4,2)(4,10)
\psarc(6,6){2}{270}{90}
\psarc(6,6){4}{270}{90}
\psdot(2,2)
\psdot(2,4)
\psdot(2,8)
\psdot(2,10)
\psdot(4,2)
\psdot(4,4)
\psdot(4,8)
\psdot(4,10)
\psdot(1,4)
\psdot(1,2)
\psdot(2,1)
\psdot(4,1)
\psarc{<-}(0,5){1}{0}{90}
\psline(1,5.2)(1,4)
\arrowLine(1,4)(1,2)
\psarc{->}(2,2){1}{180}{225}
\psarc(2,2){1}{220}{270}
\arrowLine(2,1)(4,1)
\psline(4,1)(5.2,1)
\psarc{-<}(5,0){1}{0}{90}
\psarc(5,0){1}{40}{90}
\rput(-1,10){$x_1$}
\rput(-1,8){$x_2$}
\rput(-1,6){$y_3$}
\rput(-1,4){$x_2$}
\rput(-1,2){$x_1$}
\rput(2,-1){$y_1$}
\rput(4,-1){$y_2$}
\rput(6,-1){$x_3$}
\endpspicture
\]
\caption{The proof of recursive relation (\ref{e:18})}
\label{f:arr}
\end{figure}
Note that in the middle of the process the weights of moved vertices are
not
defined by the standard rule. Actually the roles of the parameters
$x_{m+1}$ and $y_{m+1}$ interchange. In the final state the standard rule
can be used if we interchange the spectral parameters $x_{m+1}$ and
$y_{m+1}$, as it is performed in Figure \ref{f:arr}. The last graph in
Figure~\ref{f:arr} proves the statement of the lemma.
\end{proof}

The next lemma is a direct consequence of the previous one.

\begin{lemma} \label{l:19}
If $y_{m+1} = a x_{m+1}$, then
\begin{equation}
\widetilde Z_{\mathrm{HT}}(2m + 1; \bm x, \bm y) = \prod_{i=1}^m \left[ (a
y_{m+1}^2 - \bar a x_i^2)(a y_i^2 - \bar a x_{m+1}^2) \right] \widetilde
Z_{\mathrm{HT}}(2m; \bm x \smallsetminus x_{m+1}, \bm y \smallsetminus
y_{m+1}). \label{e:19}
\end{equation}
\end{lemma}

Now we are able to find the polynomial $S_\HT(2m+1, x_{m+1}, y_{m+1})$.

\begin{lemma}
The polynomial $S_\HT(2m+1, x_{m+1}, y_{m+1})$ has the form
\begin{equation}
S_\HT(2m + 1; x_{m+1}, y_{m+1}) = \left[ \prod_{i=2}^{2m} \sigma(a^i)
\right] \left(\sigma(a^{2m+1}) x_{m+1}^{2m} - \sigma(a^{2m}) x_{m+1}^{2m-1}
y_{m+1} \right). \label{e:20}
\end{equation}
\end{lemma}

\begin{proof}
Consider a state of square ice with a half-turn symmetric boundary which
gives a nontrivial contribution to the polynomial $S_\HT(2m+1; x_m, y_m)$.
It is clear that for such a state each line with the spectral parameter
$y_j$, $j = 1, \ldots, m$ may have only one vertex of first or second type,
and the same is true for each line with the spectral parameter $x_i$, $i =
1, \ldots, m$. For the line with the spectral parameters $x_{m+1}$ and
$y_{m+1}$ we have two possibilities. Either it has no vertices of first or
second type, or it has two vertices of first type. Actually in the former
case one has a hidden vertex of first type at the turning-point and in the
latter case it has a hidden vertex of second type there. One can get
convinced that the polynomial $S_\HT(2m + 1; x_{m+1}, y_{m+1})$ has the
form
\[
S_\HT(2m + 1; x_{m+1}, y_{m+1}) = C_\HT(2m+1) x_{m+1}^{2m} + D_\HT(2m+1)
x_{m+1}^{2m-1} y_{m+1}.
\]
It follows from the recursive relation (\ref{e:19}) that
\[
C_\HT(2m+1) + a D_\HT(2m+1) = a^{-2m} C_\HT(2m).
\]
Using (\ref{e:16}) one obtains the equality
\[
D_\HT(2m+1) = - \left[ \prod_{i=2}^{2m} \sigma(a^i) \right] \sigma(a^{2m})
\]
which immediately leads to (\ref{e:19}).
\end{proof}

\subsection{Main theorem}

\begin{theorem} \label{t:1}
The partition function for the square-ice model with half-turn symmetric
boundary conditions can be represented as
\begin{eqnarray}
\lefteqn{Z_\HT(2m+1; \bm x, \bm y) = \frac{a x_{m+1} y_{m+1}}{\sigma (a) (a
x_{m+1} + y_{m+1})(a y_{m+1} + x_{m+1})}} \nonumber \\*
&& \hspace{4em} {} \times \biggl[ Z(m+1; \bm x, \bm y) Z_\HT^{(2)}(2m; \bm
x \smallsetminus x_{m+1}, \bm y \smallsetminus y_{m+1}) \nonumber \\*
&& \hspace{8em} {} + Z(m; \bm x \smallsetminus x_{m+1}, \bm y
\smallsetminus y_{m+1}) Z_\HT^{(2)}(2m+2; \bm x, \bm y)
\biggr]. \label{e:21}
\end{eqnarray}
\end{theorem}

\begin{proof}
Consider first the modified partition function $\widetilde Z_\HT(2m+1; \bm
x, \bm y)$. The comparision of recursive relations (\ref{e:15}) and
(\ref{e:19}) with recursive relations (\ref{e:2}) and (\ref{e:10}) suggests
to
use for it the following ansatz
\begin{eqnarray*}
\lefteqn{\widetilde Z_\HT(2m+1; \bm x, \bm y)} \\
&& \hspace{4em} {} = A(x_{m+1}, y_{m+1}) \widetilde Z(m+1; \bm x, \bm y)
\widetilde Z_\HT^{(2)}(2m; \bm x \smallsetminus x_{m+1}, \bm y
\smallsetminus y_{m+1}) \\
&& \hspace{5em} {} + B(x_{m+1}, y_{m+1}) \widetilde Z(m; \bm x
\smallsetminus x_{m+1}, \bm y \smallsetminus y_{m+1}) \widetilde
Z_\HT^{(2)}(2m+2; \bm x, \bm y).
\end{eqnarray*}
This ansatz satisfies recursive relation (\ref{e:15}) provided that the
functional form of the coefficients $A(x_{m+1}, y_{m+1})$ and $B(x_{m+1},
y_{m+1})$ is the same for all $m$. Comparing the terms of the maximal total
degree in all spectral parameters except $x_{m+1}$ and $y_{m+1}$ in both
sides of the above equality, and using relations (\ref{e:20}), (\ref{e:5}),
(\ref{e:13}), (\ref{e:4}) and (\ref{e:12}), one comes to the condition
\begin{eqnarray*}
\lefteqn{\frac{1}{\sigma(a)} \left[ \sigma(a^{2m+1}) x_{m+1}^2 -
\sigma(a^{2m}) x_{m+1} y_{m+1} \right]} \\
&& \hspace{2em} {} = A(x_{m+1}, y_{m+1}) \left[ \sigma(a^{2m+2}) x_{m+1}^2
- \sigma(a^{2m}) y_{m+1}^2 \right] \\[.5em]
&& \hspace{3em} {} + B(x_{m+1}, y_{m+1}) \, x_{m+1}^2 \left[
\sigma(a^{2m+1}) x_{m+1}^2 - \sigma(a^{2m-1}) y_{m+1}^2 \right]
\end{eqnarray*}
One can get convinced that this condition is satisfied for each $m$ if and
only if
\begin{eqnarray*}
&& A(x_m, y_m) = \frac{a x_{m+1} y_{m+1}}{\sigma (a) (a x_{m+1} +
y_{m+1})(a y_{m+1} + x_{m+1})}, \\[.5em]
&& B(x_m, y_m) = \frac{a}{\sigma (a) (a x_{m+1} + y_{m+1})(a y_{m+1} +
x_{m+1})}.
\end{eqnarray*}
Taking into account Lemma \ref{l:17}, we conclude that we obtain a right
expression for the modified partition function $\widetilde Z_\HT(2m+1; \bm
x, \bm y)$. Moving to functions without tildes we see that
relation~(\ref{e:21}) is true.
\end{proof}

\subsection{Two types of half-turn symmetric alternating-sign matrices}

As we remarked above a half-turn symmetric alternating-sign matrix of odd
order has either~$1$ or~$-1$ as its central entry. It appears that one can
separate contribution of these two types of the matrices into the partition
function. To this end note first that the weight of the vertices
corresponding to~$1$ or~$-1$ entries of an alternating-sign matrix is an
even function of the parameter $a$, while the weights of the vertices
corresponding to 0 entries are odd functions of $a$. Let the number of $-1$
entries in an $n \times n$ alternating-sign matrix is equal to $k$, then
the number of $1$ entires in this matrix is equal to $n+k$. Hence, the
number of its $0$ entries is $n(n-1) - 2k$. This number is always even and
we have 
\[
Z(m; \bm x, \bm y)|_{a \to -a} = Z(m; \bm x, \bm y).
\]
A similar consideration for the case of half-turn symmetric
alternating-sign matrices of even order leads to the conclusion that
\begin{equation}
Z_\HT(2 m; \bm x, \bm y)|_{a \to -a} = (-1)^m Z_\HT(m; \bm x, \bm y).
\label{e:22}
\end{equation}
This equality, in particular, gives
\begin{equation}
Z_\HT^{(2)}(2m; \bm x, \bm y)|_{a \to -a}=(-1)^m Z_\HT^{(2)}(2m, \bm x, \bm
y). \label{e:23}
\end{equation}

Consider now a half-turn symmetric alternating-sign matrix of odd order
$(2m+1) \times (2m+1)$. Let the central entry of the matrix is $1$. In this
case the number of $-1$ entries is even, say $2l$. Then the number of $0$
entries is $(2m+1)^2 - (2m + 1) - 4l = 2m(2m+1) - 4l$. Only a half of these
vertices corresponds to vertices of the corresponding graph, describing the
state of square-ice with a half-turn symmetric boundary. Therefore, the
weight of the state under consideration acquires the factor $(-1)^m$ under
the replacement $a \to -a$. In a similar way we can see that the weight of
the state corresponding to a half-turn symmetric alternating-sign matrix of
odd order whose central entry is $-1$ acquires the factor $(-1)^{m+1}$
under this replacement. Thus, representing $Z_\HT (2m+1; \bm x, \bm y)$ in
the
form
\[
Z_\HT(2m+1; \bm x, \bm y) = Z_\HT^{(+)} (2m+1; \bm x, \bm y) +
Z_\HT^{(-)}(2m+1; \bm x, \bm y),
\]
where
\begin{eqnarray*}
& Z_\HT^{(+)} (2m+1; \bm x, \bm y)|_{a \to -a} = (-1)^m Z_\HT^{(+)} (2m+1;
\bm x, \bm y), \\[.5em]
& Z_\HT^{(-)} (2m+1; \bm x, \bm y)|_{a \to -a} = (-1)^{m+1} Z_\HT^{(-)}
(2m+1; \bm x, \bm y),
\end{eqnarray*}
we separate the contribution of two types of the matrices. It is clear that
\begin{eqnarray*}
&& Z^{(+)}_\HT(2m+1; \bm x, \bm y) = \frac{1}{2} \left[ Z_\HT(2m+1; \bm x,
\bm y) + (-1)^m Z_\HT(2m+1; \bm x, \bm y) |_{a \to -a}
\right], \\
&& Z^{(-)}_\HT(2m+1; \bm x, \bm y) = \frac{1}{2} \left[ Z_\HT(2m+1; \bm x,
\bm y) - (-1)^m Z_\HT(2m+1; \bm x, \bm y)|_{a \to -a} \right].
\end{eqnarray*}
Using relation (\ref{e:21}) and taking into account equalities (\ref{e:22})
and (\ref{e:23}), one obtains the following theorem.
\begin{theorem} \label{t:2}
The contributions to the partition function $Z\HT(2m+1; \bm x, \bm y)$ of
the states corresponding to the alternating-sign matrices having~$1$
or~$-1$ as the central entry are
\begin{eqnarray}
\lefteqn{Z^{(+)}_\HT(2m+1; \bm x, \bm y) = \frac{1}{\sigma (a) \sigma(a
x_{m+1} \bar y_{m+1}) \sigma(a \bar x_{m+1} y_{m+1}) }} \nonumber \\*
&& {} \times \biggl[ (a + \bar a) Z(m+1; \bm x, \bm y) Z_\HT^{(2)}(2m; \bm
x \smallsetminus x_{m+1}, \bm y \smallsetminus y_{m+1}) \nonumber \\*
&& \hspace{1em} {} - (x_{m+1} \bar y_{m+1} + \bar x_{m+1} y_{m+1}) Z(m; \bm
x \smallsetminus x_{m+1}, \bm y \smallsetminus y_{m+1}) Z_\HT^{(2)}(2m+2;
\bm x, \bm y) \biggr], \label{e:24} \\
\lefteqn{Z^{(-)}_\HT(2m+1; \bm x, \bm y) = \frac{1}{\sigma (a) \sigma(a
x_{m+1} \bar y_{m+1}) \sigma(a \bar x_{m+1} y_{m+1}) }} \nonumber \\*
&& {} \times \biggl[ - (x_{m+1} \bar y_{m+1} + \bar x_{m+1} y_{m+1}) Z(m+1;
\bm x, \bm y) Z_\HT^{(2)}(2m; \bm x \smallsetminus x_{m+1}, \bm y
\smallsetminus y_{m+1}) \nonumber \\*
&& \hspace{8em} {} + (a + \bar a) Z(m; \bm
x \smallsetminus x_{m+1}, \bm y \smallsetminus y_{m+1}) Z_\HT^{(2)}(2m+1;
\bm x, \bm y) \biggr] \label{e:25}
\end{eqnarray}
respectively.
\end{theorem}

\section{Special determinant representations}

It appears that in the case $a=\rme^{\rmi \pi/3}$ partition functions of
many square-ice models possess additional symmetry properties and new
representations. It is interesting to consider from this point of view the
square-ice model with half-turn symmetric boundary condition. Assume
that $a=\rme^{\rmi \pi/3}$ and discuss first additional properties of the
partition functions $Z(n; \bm x, \bm y)$ and $Z^{(2)}_\HT(2m, \bm x, \bm
y)$ arising in this case. 

Consider the partition function $Z(n; \bm x, \bm y)$ as a function of the
$2n$-dimensional vector $\bm u = (u_1, \ldots, u_{2n})$, where $u_{2i-1} =
x_i$ and $u_{2i} = y_i$. From recursive relation (\ref{e:3}) one obtains
that if $u_{2n} = a u_{2n-1}$, then
\begin{equation}
Z(n; \bm u) = \sigma(a) \prod_{\mu=1}^{2n-2} \sigma(a u_\mu \bar u_{2n-1})
Z(n-1; \bm u {\smallsetminus} u_{2n-1} {\smallsetminus}
u_{2n}). \label{e:26}
\end{equation}
It can be shown that for any $\mu = 1, \ldots, 2n$ the partition
function $Z(n; \bm u)$ satisfies also the relation
\begin{eqnarray}
&& Z(n; (u_1, \ldots, u_\mu, \ldots, u_{2n})) \prod_{\nu \ne \mu}
\sigma(u_\nu \bar u_\mu) \nonumber \\
&& \hspace{4em} {} + Z(n; (u_1, \ldots, a^2 u_\mu, \ldots, u_{2n}))
\prod_{\nu \ne \mu} \sigma(u_\nu \bar a^2 \bar u_\mu) \nonumber \\
&& \hspace{8em} {} + Z(n; (u_1, \ldots, \bar a^2 u_\mu, \ldots, u_{2n}))
\prod_{\nu \ne \mu} \sigma (u_\nu a^2 \bar u_\mu) = 0.
\label{e:27}
\end{eqnarray}
These relations allow one to obtain the following determinant
representation for the partition function $Z(n, \bm u)$:
\begin{equation}
Z(n; \bm u) =  (-1)^{n(n-1)/2} \frac{\sigma^n(a)}{\prod_{\mu < \nu}
\sigma(u_\mu \bar u_\nu)} \det P(n, \bm u),  \label{e:28}
\end{equation}
where
\[
P(n; \bm u) = \left( \begin{array}{ccccc}
u_1^{3n-2}&u_2^{3n-2}&u_3^{3n-2}&\dots&u_{2n}^{3n-2} \\
u_1^{3n-4}&u_2^{3n-4}&u_3^{3n-4}&\dots&u_{2n}^{3n-4}  \\
u_1^{3n-8}&u_2^{3n-8}&u_3^{3n-8}&\dots&u_{2n}^{3n-8}  \\
\vdots&\vdots&\vdots&\ddots&\vdots   \\
u_1^{-3n+2}&u_2^{-3n+2}&u_3^{-3n+2}&\dots&u_{2n}^{-3n+2}
\end{array} \right),
\]
see~\cite{Str02, RazStr04}. It follows from the above representation that
the
partition function $Z(n; \bm u)$ is symmetric in the coordinates of the
vector $\bm u$.

For the function $Z^{(2)}_\HT(2m; \bm x, \bm y)$ considered
as a function of the $2m$-dimensional vector $\bm u = (u_1, \ldots,
u_{2m})$, at $a=\rme^{\rmi \pi/3}$ one obtains actually the same relations
\begin{equation}
Z^{(2)}_\HT(2m; \bm u) = \sigma(a) \prod_{\mu=1}^{2m-2} \sigma(a \, u_\mu
\, \bar u_{2m-1}) Z^{(2)}_\HT(2m-2; \bm u {\smallsetminus} u_{2m-1}
{\smallsetminus} u_{2m}) \label{e:29}
\end{equation}
and
\begin{eqnarray}
&& Z^{(2)}_\HT(2m; (u_1, \ldots, u_\mu, \ldots, u_{2m})) \prod_{\nu \ne
\mu} \sigma(u_\nu \bar u_\mu) \nonumber \\*
&& \hspace{4em} {} + Z^{(2)}_\HT(2m; (u_1, \ldots, a^2 u_\mu, \ldots,
u_{2m})) \prod_{\nu \ne \mu} \sigma(u_\nu \bar a^2 \bar u_\mu) \nonumber
\\*
&& \hspace{8em} {} + Z^{(2)}_\HT(2m; (u_1, \ldots, \bar a^2 u_\mu, \ldots,
u_{2m})) \prod_{\nu \ne \mu} \sigma (u_\nu a^2 \bar u_\mu) = 0,
\label{e:30}
\end{eqnarray}
as for the partiton function $Z(n, \bm u)$. They give the determinant
representation \cite{Str04}
\begin{equation}
Z^{(2)}_\HT(2m; \bm u) =  (-1)^{m(m-1)/2} \frac{\sigma^m(a)}{\prod_{\mu <
\nu} \sigma(u_\mu \bar u_\nu)} \det Q(m, \bm u),  \label{e:31}
\end{equation}
where
\[
Q(m; \bm u) = \left( \begin{array}{ccccc}
u_1^{3m-1}&u_2^{3m-1}&u_3^{3m-1}&\dots&u_{2m}^{3m-1} \\
u_1^{3m-5}&u_2^{3m-5}&u_3^{3m-5}&\dots&u_{2m}^{3m-5}  \\
u_1^{3m-7}&u_2^{3m-7}&u_3^{3m-7}&\dots&u_{2m}^{3m-7}  \\
\vdots&\vdots&\vdots&\ddots&\vdots   \\
u_1^{-3m+1}&u_2^{-3m+1}&u_3^{-3m+1}&\dots&u_{2m}^{-3m+1}
\end{array} \right).
\]
We again have symmetricity in the coordinates of the vector $\bm u$.
Therefore, the partition function $Z_\HT(2m, \bm x, \bm y)$ at $a =
\rme^{\rmi \pi/3}$ is symmetric in the union of the coordinates of the
vectors~$\bm x$ and~$\bm y$.

From the other hand, the partition function $Z_\HT(2m+1; \bm x, \bm y)$ at
$a = \rme^{\rmi \pi/3}$  is not symmetric in the union of the coordinates
of the vectors $\bm x$ and $\bm y$. Assume again that $a=\rme^{\rmi \pi/3}$
and consider the function
\[
Z'_\HT(2m+1; \bm u) = Z_\HT(2m+1; (u_1, u_3, \ldots, u_{2m-1}, u_{2m+1}),
(u_2, u_4, \ldots, u_{2m}, u_{2m+1})),
\]
where $\bm u = (u_1, \ldots, u_{2m+1})$. As it follows from the next
theorem the function $Z'_\HT(2m+1; \bm u)$ is symmetric in the coordinates
of the vector $\bm u$.

\begin{theorem} \label{t:3}
The function $Z'_\HT(2m+1; \bm u)$ has the following determinant
representation
\begin{equation}
Z'_\HT(2m+1, \bm u) = \frac{\sigma^{2m}(a)}{\displaystyle \prod_{\mu < \nu}
\sigma^2(u_\mu \bar u_\nu)} \det P'(m+1; \bm u) \det P'(m+1; \bar{\bm u}),
\label{e:32}
\end{equation}
where $P'(m, \bm u)$ is the $(2m-1)\times(2m-1)$ matrix which is obtained
from $P(m, \bm u)$ when one removes the last column and the last row.
\end{theorem}

\begin{proof}
Equality (\ref{e:21}) implies
\begin{eqnarray}
\lefteqn{Z'_\HT(2m+1; \bm u) = - \frac{1}{\sigma^3(a)}} \nonumber \\
&& \hspace{1em} {} \times \biggl[Z(m+1; (u_1, \ldots, u_{2m}, u_{2m+1},
u_{2m+1})) Z^{(2)}_\HT(2 m; (u_1, \ldots, u_{2m})) \nonumber \\
&& \hspace{3em} {} + Z^{(2)}_\HT(2m+2; (u_1, \ldots, u_{2m}, u_{2m+1},
u_{2m+1})) Z(m; (u_1, \ldots, u_{2m})) \biggr]. \label{e:33}
\end{eqnarray}
Let us multiply equation (\ref{e:27}) for $n=m$ and $\mu=2m$ by $Z_\HT(m;
(u_1, \ldots, \bar a^2 u_{2m}))$, equation (\ref{e:30}) for $\mu=2m$ by
$Z(m; (u_1, \ldots, \bar a^2 u_{2m}))$ and find the difference of the
obtained expressions. The result can be written as
\begin{equation}
\frac{W(m; (u_1, \ldots, u_{2m-1}, u_{2m}))}{\prod_{\nu =
1}^{2m-1}\sigma(u_\nu \bar u_{2m})} = \frac{W(m; (u_1, \ldots, u_{2m-1},
a^2 u_{2m}))}{\prod_{\nu = 1}^{2m-1}\sigma(u_\nu \bar a^2 \bar u_{2m})},
\label{e:34}
\end{equation}
where we introduced the `Wronskian'
\begin{eqnarray*}
\lefteqn{W(m; \bm u) = Z(m; (u_1, \ldots, u_{2m-1}, \bar a^2 u_{2m}))
Z^{(2)}_\HT(2m; (u_1, \ldots, u_{2m-1}, a^2 u_{2m}))} \\
&& \hspace{4em} {} - Z^{(2)}_\HT(2m; (u_1, \ldots, u_{2m-1}, \bar a^2
u_{2m})) Z(m; (u_1, \ldots, u_{2m-1}, a^2 u_{2m})).
\end{eqnarray*}
The function $W(m; \bm u)$ is a centered Laurent polynomial in $u_{2m}$ of
width $2m-1$. The product $\prod_{\mu=1}^{2m-1}\sigma(u_\mu \bar u_{2m})$
is also a centered Laurent polynomial of the same width. Multiplying the
nominators and denominators of the fractions in both sides of equality
(\ref{e:34}) by $u_{2m}^{2m-1}$, we see that these fractions are rational
functions of $u_{2m}^2$. The positions of possible poles in $u_{2m}^2$ of
the left-hand and right-hand sides of (\ref{e:34}) are different. This
means that the equality can be true only if the rational functions under
consideration are constant in $u_{2m}^2$. Thus, one has
\begin{equation}
W(m; \bm u) = w(m; \bm u {\smallsetminus} u_{2m}) \prod_{\mu=1}^{2m-1}
\sigma(u_\mu \bar u_{2m}). \label{e:35}
\end{equation}
Using the equalities
\begin{eqnarray*}
& Z(m; (u_1, \ldots, -u_\mu, \ldots, u_{2m})) = (-1)^{m-1} Z(m; (u_1,
\ldots, u_\mu, \ldots, u_{2m})), \\
& Z^{(2)}_\HT(m; (u_1, \ldots, -u_\mu, \ldots, u_{2m})) = (-1)^m
Z^{(2)}_\HT(m; (u_1, \ldots, u_\mu, \ldots, u_{2m})),
\end{eqnarray*}
and recursive relations (\ref{e:26}) and (\ref{e:29}), one can get
convinced
that
\begin{eqnarray*}
\lefteqn{W(m; (u_1, \ldots, u_{2m-2}, u_{2m-1}, a^2 u_{2m-1})) = (-1)^m
\sigma (a) \prod_{\mu=1}^{2m-2} \sigma(a u_\mu \bar u_{2m-1})} \\
&& \hspace{1em} {} \times \biggl[Z(m; (u_1, \ldots, u_{2m_2}, u_{2m-1},
u_{2m-1})) Z^{(2)}_\HT(2m-2; (u_1, \ldots, u_{2m-2})) \\
&& \hspace{3em} {} + Z^{(2)}_\HT(2m; (u_1, \ldots, u_{2m-2}, u_{2m-1},
u_{2m-1})) Z(m-1; (u_1, \ldots, u_{2m-2})) \biggr].
\end{eqnarray*}
From the other hand, it follows from (\ref{e:35}) that
\begin{equation}
W(m; (u_1, \ldots, u_{2m-1}, a^2 u_{2m-1})) = - \sigma(a)
\prod_{\mu=1}^{2m-2} \sigma(a u_\mu \bar u_{2m-1}) w(m, (u_1, \ldots,
u_{2m-1})). \label{e:36}
\end{equation}
The above two equalities and (\ref{e:33}) give
\begin{equation}
Z_\HT(2m+1; \bm u) = \frac{(-1)^{m+1}}{\sigma^3(a)} w(m+1; \bm u).
\label{e:37}
\end{equation}
From the determinant representations (\ref{e:28}) and (\ref{e:31}) one
obtains that as $u_{2m} \to 0$
\begin{eqnarray*}
& \displaystyle Z(m, \bm u) \sim \frac{1}{u_{2m}^{m-1}}
\frac{(-1)^{m(m-1)/2} \sigma^m(a)}{\displaystyle \prod_{\mu=1}^{2m-1} u_\mu
\prod_{\scriptstyle \mu,\nu = 1 \atop \scriptstyle \mu < \nu}^{2m-1}
\sigma(u_\mu \bar u_\nu)} \det P'(m; \bm u {\smallsetminus} u_{2m}) +
\ldots, \\
& \displaystyle Z^{(2)}_\HT(2m, \bm u) \sim \frac{1}{u_{2m}^m}
\frac{(-1)^{m(m-1)/2}
\sigma^m(a)}{\displaystyle \prod_{\mu=1}^{2m-1} u_\mu \prod_{\scriptstyle
\mu,\nu = 1 \atop \scriptstyle \mu < \nu}^{2m-1} \sigma(u_\mu \bar u_\nu)}
\det Q'(m; \bm u {\smallsetminus} u_{2m}) + \ldots,
\end{eqnarray*}
where $Q'(m, \bm u)$ is the $(2m-1)\times(2m-1)$ matrix which is obtained
from $Q(m, \bm u)$ when one removes the last column and the last row. These
relations show that
\[
W(m, \bm u) \sim - \frac{1}{u_{2m}^{2m-1}}
\frac{\sigma^{2m+1}(a)}{\displaystyle \prod_{\mu=1}^{2m-1} u_\mu^2
\prod_{\scriptstyle \mu,\nu = 1 \atop \scriptstyle \mu < \nu}^{2m-1}
\sigma^2(u_\mu \bar u_\nu)} \det P'(m; \bm u {\smallsetminus} u_{2m})
\det Q'(m; \bm u {\smallsetminus} u_{2m}) + \ldots
\]
as $u_{2m} \to 0$. Using the equality
\[
\det P'(m, \bar{\bm u}) = (-1)^m \prod_{\mu=1}^{2m-1} u_\mu^{-3} \det Q'(m,
\bm u)
\]
and relation (\ref{e:36}), we obtain
\[
w(m, \bm u) = (-1)^m \frac{\sigma^{2m+1}(a)}{\displaystyle \prod_{\mu <
\nu} \sigma^2(u_\mu \bar u_\nu)} \det P'(m; \bm u) \det P'(m; \bar{\bm u}).
\]
The statement of the lemma follows now from (\ref{e:37}).
\end{proof}

Using quite different technique, Okada also obtained the determinant
representations (\ref{e:28}) and (\ref{e:31}) \cite{Oka04}. He expressed
the results in terms of characters of classical groups and conjectured, in
particular, that the number the half-turn symmetric alternating-sign
matrices of odd order is connected with the dimension of some specific
representation of $\mathrm{GL}(2n+1)\times \mathrm{GL}(2n+1)$. Possible
generalizations of this conjecture have been discussed by
Kuperberg~\cite{Kup04}. We hope that our determinant representation
(\ref{e:32}) sheds a new light on this question.

\section{Enumeration results}

\subsection{Refined $\bm x$-enumerations}

Denote by $A(n; x)$ the total weight of the $n \times n$ alternating-sign
matrices, where the weight of an individual alternating-sign matrix is
$x^k$ if it has $k$ matrix elements equal to $-1$. The quantities $A(n; x)$
are called $x$-enumerations of the alternating-sign matrices.\footnote{Do
not mix the vector $\bm x$ and the parameter of the enumeration $x$.}
Considering the partition function $Z(n; \bm x, \bm y)$ at
$\bm x = \bm y = \bm 1$, where $\bm 1 = (1, \ldots, 1)$, one obtains
information on $x$-enumerations. Namely, one has the equality
\[
A(n; x) = \frac{1}{\sigma^{n^2-n}(a) \sigma^n(a^2)} Z(n; \bm 1, \bm 1),
\]
where
\[
x = \left[ \frac{\sigma(a^2)}{\sigma(a)} \right]^2 = (a + \bar a)^2.
\]
In particular, if $a = \rme^{\rmi \pi/3}$, then $x = 1$, and the above
equality gives the total number of the $n \times n$ alternating-sign
matrices $A(n) = A(n, 1)$. In paper \cite{Kup96}, Kuperberg used the
Izergin--Korepin determinant representation for the partition function
$Z(n; \bm x, \bm y)$ \cite{Ize87, KorIzeBog93} and proved the formula for
$A(n)$ conjectured by Mills, Robbins and Rumsey \cite{MilRobRum82,
MilRobRum83} and first proved by Zeilberger \cite{Zei96a}.

One also defines refined $x$-enumerations of the alternating-sign matrices.
Recall that an alternating-sign matrix has only one entry 1 in the first
column, all other entries are zero. Denote by $A(n, r; x)$ the total weight
of the $n \times n$ alternating-sign matrices having 1 at the $r$th
position of the first column. Here the weight of an individual
alternating-sign matrix is $x^k$ if it has $k$ matrix elements equal to
$-1$. It is readily seen that we have the equality
\begin{equation}
\mathcal A(n; t, x) \equiv \sum_{r=1}^{n} A(n, r; x) t^{r-1} = 
\frac{Z(n; \bm 1, (v, 1, \ldots, 1))} {\sigma^{n^2-2n+1}(a)
\sigma^n(a^2) \sigma^{n-1}(a v)}, \label{e:38}
\end{equation}
where
\[
x = \left[ \displaystyle \frac{\sigma(a^2)}{\sigma(a)} \right]^2, \qquad t
= \frac{\sigma(a \bar v)}{\sigma(a v)}.
\]
If $a = \rme^{\rmi \pi/3}$ one obtains formulas for refined enumerations of
the alternating-sign matrices $A(n, r) = A(n, r; 1)$. In paper
\cite{Zei96b}, Zeilberger used the Izergin--Korepin determinant
representation for $Z(n; \bm x, \bm y)$ to prove the refined
alternating-sign matrix conjecture for $A(n, r)$ by Mills, Robbins,
and Rumsey~\cite{MilRobRum82, MilRobRum83}.

For the case of half-turn symmetric alternating-sign matrices one has
\[
{\mathcal A}_\HT(2m; t,x) \equiv \sum_{r=1}^{2m} A_\HT(2m, r; x) t^{r-1}
= \frac{Z_\HT(2m; \bm 1, (v, 1, \ldots, 1))} {\sigma^{2m^2-3m+1}(a)
\sigma^m(a^2) \sigma^{2m-1}(a v)}.
\]
Note that here the weight of an individual alternating-sign matrix is
$x^{k/2}$ if it has $k$ entries equal to $-1$. Since in the case under
consideration the number of $-1$ entries is always even, we weigh in
accordance with the number of symmetry orbits of $-1$ entries. It is
convenient to introduce the notation
\begin{equation}
\mathcal A_\HT^{(2)}(2m; t, x) \equiv \frac{\mathcal A_\HT(2m; t,
x)}{\mathcal A(m; t, x)} = \frac{Z_\HT^{(2)}(2m; \bm 1, (v, 1, \ldots,
1))} {\sigma^{m^2-m}(a) \sigma^m(a v)}. \label{e:39}
\end{equation}
Further, for the case of the half-turn symmetric alternating-sign matrices
of odd order we obtain
\begin{equation}
\mathcal A_\HT(2m+1; t,x) \equiv \sum_{r=1}^{2m+1} A_\HT(2m+1, r; x)
t^{r-1} = \frac{Z_\HT(2m+1; \bm 1, (v, 1, \ldots, 1))} 
{\sigma^{2m^2-m}(a) \sigma^m(a^2) \sigma^{2m}(a v)}. \label{e:40}
\end{equation}
Here again the weight of an individual alternating-sign matrix is $x^{k/2}$
if it has $k$ entries equal to~$-1$. Note that the number of $-1$ entries
is even or odd, if the central entry of the matrix is~$1$ or $-1$
respectively. Robbins used the weighing in accordance with the number of
symmetry orbits of $-1$ entries in this case as well \cite{Rob00}. Our
definition seems more convenient. The connection with the $x$-enumeration
used by Robbins is given below.

Having in mind representation (\ref{e:21}), from (\ref{e:40}), (\ref{e:38})
and (\ref{e:39}) we obtain
\begin{equation}
\mathcal A_\HT(2m+1; t, x) = \frac{\sqrt{x} \mathcal A(m+1; t, x) \mathcal
A_\HT^{(2)}(2m; t, x) + \mathcal A(m; t, x) \mathcal A_\HT^{(2)}(2m+2;
t,x)}{\sqrt{x} + 2}. \label{e:41}
\end{equation}
While $\mathcal A(m; t, x)$ and $\mathcal A_\HT^{(2)}(m; t, x)$ are
polynomials in the variables $x$ and $t$, $\mathcal A_\HT(2m+1; t, x)$ has
also half-integer powers of the variable $x$. One can separate it into two
parts
\[
\mathcal A_\HT(2m+1; t, x) = \mathcal A_\HT^{(+)}(2m+1; t, x) + \sqrt{x}
\mathcal A_\HT^{(-)}(2m+1; t, x),
\]
where $\mathcal A_\HT^{(+)}(2m+1, t, x)$ and $\mathcal A_\HT^{(-)}(2m+1; t,
x)$ are polynomials in the variable $x$. These two parts give the refined
$x$-enumerations of the half-turn symmetric alternating-sign matrices of
odd order with~$1$ and~$-1$ in the centre of a matrix respectively. 
Relation (\ref{e:41}) gives\footnote{Certainly, one can directly use
relations (\ref{e:24}) and (\ref{e:25}).}
\begin{eqnarray}
\lefteqn{\mathcal A_\HT^{(+)}(2m+1; t, x)} \nonumber \\
&& \hspace{2em} {} =  \frac{{} -x \mathcal A(m+1; t, x)
\mathcal A_\HT^{(2)}(2m; t, x) + 2 \mathcal A(m; t, x) \mathcal
A_\HT^{(2)}(2m+2; t, x)}{4-x}, \label{e:42} \\
\lefteqn{\mathcal A_\HT^{(-)}(2m+1; t, x)} \nonumber \\
&& \hspace{4em} {} = \frac{2 \mathcal A(m+1; t, x) \mathcal A_\HT^{(2)}(2m;
t, x) - \mathcal A(m; t, x) \mathcal A_\HT^{(2)}(2m+2; t, x)}{4-x}.
\label{e:43}
\end{eqnarray}

Robbins used the $x$-enumeration of the half-turn symmetric
alternating-sign matrices related to the number of symmetry orbits of $-1$
entries \cite{Rob00}. It is clear that such $x$-enumeration has the form
\begin{eqnarray*}
\lefteqn{\mathcal A_\HT^{\mathrm R}(2m+1; t, x) = \mathcal
A_\HT^{(+)}(2m+1; t, x) + x \mathcal A_\HT^{(-)}(2m+1; t, x)} \\[.5em]
&& {} = \frac{x \mathcal A(m+1; t, x) \mathcal A_\HT^{(2)}(2m, t, x) +
(2-x) \mathcal A(m; t, x) \mathcal A_\HT^{(2)}(2m+2; t, x)}{4-x}.
\end{eqnarray*}

We see that the refined $x$-enumerations $\mathcal A_\HT^{(+)}(2m+1; t, x)$
and $\mathcal A_\HT^{(-)}(2m+1; t, x)$ are determined by the polynomials
$\mathcal A(m; t, x)$ and $\mathcal A_\HT^{(2)}(2m; t, x)$ (or $\mathcal
A_\HT(2m; t, x)$). These polynomials for general $x$ are not known yet.
However, for $x=1$ we know the explicit form of these polynomials. This
allows us to obtain the expressions for the refined 1-enumerations of the
haft-turn symmetric alternating-sign matrices of odd order. We start with
the usual enumerations (1-enumerations).

\subsection{1-enumerations}

Recall that the total number of the alternating-sign matrices is given by
the formula
\[
A(m) = \prod_{i=0}^{m-1} \frac{(3i+1)!}{(m+i)!},
\]
and for the total number of half-turn symmetric alternating-sign matrices
one has
\[
A_\HT(2m) = \prod_{i=0}^{m-1} \frac{(3i)! \, (3i+2)!}{[(m+i)!]^2}.
\]
It follows from these relations that
\begin{eqnarray}
& \displaystyle \frac{A(m+1)}{A(m)} = \frac{m!\,(3m+1)!}{(2m)! \, (2m+1)!},
\label{e:44} \\[.5em]
& \displaystyle \frac{A_\HT(2m+2)}{A_\HT(2m)} = \frac{[m!]^2 \, (3m)! \,
(3m+2)!}{[(2m)! \, (2m+1)!]^2}. \label{e:45}
\end{eqnarray}

Putting $x=1$ and $t=1$ and having in mind that $\mathcal A(n; 1, 1) =
A(n)$ and $\mathcal A_\HT^{(2)}(2m; 1, 1) = A_\HT(2m)/A(m)$, we obtain from 
(\ref{e:42}) and (\ref{e:43}) that
\begin{eqnarray*}
&& \frac{A_\HT^{(+)}(2m+1)}{A_\HT(2m)} = -\frac{1}{3} \frac{A(m+1)}{A(m)} +
\frac{2}{3} \frac{A_\HT(2m+2)}{A_\HT(2m)} \frac{A(m)}{A(m+1)}, \\
&& \frac{A_\HT^{(-)}(2m+1)}{A_\HT(2m)} = \frac{2}{3} \frac{A(m+1)}{A(m)} -
\frac{1}{3} \frac{A_\HT(2m+2)}{A_\HT(2m)} \frac{A(m)}{A(m+1)}.
\end{eqnarray*}
Using relations (\ref{e:44}) and (\ref{e:45}), one can express
$A_\HT^{(+)}(2m+1)$ and $A_\HT^{(-)}(2m+1)$ through $A_\HT(2m)$. It is
convenient to write the answer in the following form
\begin{eqnarray*}
& \displaystyle A_\HT(2m+1) = \frac{(m)! \, (3m)!}{[(2m)!]^2}
A_\HT(2m), \\
& \displaystyle A_\HT^{(+)}(2m+1) = \frac{m+1}{2m+1} A_\HT(2m+1), \qquad
A_\HT^{(-)}(2m+1) = \frac{m}{2m+1} A_\HT(2m+1).
\end{eqnarray*}
The first equality was conjectured by Robbins~\cite{Rob00} as well, but, as
far as we know, has not been proved yet.

The simplicity of the relation
\[
\frac{A_\HT^{(+)}(2m+1)}{A_\HT^{(-)}(2m+1)} = \frac{m+1}{m}
\]
is rather unexpected.

\subsection{Refined 1-enumerations}

The polynomial $\mathcal A(m; t) \equiv \mathcal A(m; t, 1)$ is determined
by the celebrated refined enumeration of the alternating-sign matrices
conjectured by Mills, Robbins, and Rumsey~\cite{MilRobRum82, MilRobRum83}
and proved by Zeilberger~\cite{Zei96b}. It has the form
\[
\frac{\mathcal A(m, t)}{A(m)} = \frac{(2m-1)!}{(m-1)! (3m-2)!} \sum_{r =
1}^m \frac{(m+r-2)! (2m-r-1)!}{ (r-1)! (m-r)!} \, t^{r-1}.
\]
The polynomial $\mathcal A_\HT^{(2)}(m; t) \equiv \mathcal A_\HT^{(2)}(m;
t, 1)$ was found in paper~\cite{Str04}. It is given by the formula
\begin{eqnarray*}
\lefteqn{\frac{\mathcal A_\HT^{(2)}(2m; t)}{A_\HT^{(2)}(2m)}} \\
&& {} = \frac{(3m-2) (2m-1)!}{(m-1)!(3m-1)!} \sum_{r=1}^{m+1}
\frac{(m^2 -  m r + (r - 1)^2)(m + r - 3)!(2 m - r - 1) 
}{(r - 1)!(m - r + 1)!} \, t^{r-1}.
\end{eqnarray*}
Using the above relations in (\ref{e:42}) and (\ref{e:43}) with $x=1$, we
obtain the refined enumerations of the half-turn symmetric alternating-sign
matrices with~$1$ and $-1$ in the centre of a matrix respectively.

\subsection{Refined 4-enumerations}

Let us return again to relations (\ref{e:42}) and (\ref{e:43}). The
denominators of the fractions in the right-hand sides of these relation
are $x-4$. From the other hand, the expression in the  right-hand side is
regular for any $x$. Therefore, one should have the equality
\[
2 \, \mathcal A(m+1; t,4) \mathcal A_\HT^{(2)}(2m; t,4) = \mathcal
A(n; t, 4) \mathcal A_\HT^{(2)}(2m+2; t, 4),
\]
which implies the recursive relation
\[
2 \, \frac{\mathcal A^2(m+1; t, 4)}{\mathcal A^2(m; t, 4)} = \frac{\mathcal
A_\HT(2m+2; t, 4)}{\mathcal A_\HT(2m; t, 4)}.
\]
Using the equalities $\mathcal A_\HT(2; t, x) = 1 + t$ and $\mathcal A(1;
t, x) = 1$, one comes to the relation
\[
\mathcal A_\HT(2m; t, 4) = 2^{m-1} (1+t) \mathcal A^2(m; t, 4).
\]

{\bf Acknowledgments}.
The work was supported in part by the Russian Foundation for Basic Research
under grant \# 04--01--00352. We are grateful to G. Kuperberg for the
interesting and stimulating correspondence.

\end{document}